\documentclass[11pt,english]{article}
\pdfoutput=1
\usepackage{jheppub}
\usepackage{lmodern}
\usepackage{lscape}
\usepackage{amsmath}
\usepackage{slashed}
\usepackage{mathtools}
\usepackage{cleveref}
\usepackage{tcolorbox}
\tcbuselibrary{breakable}
\usepackage{tikz-cd}
\usepackage{subcaption}
\usepackage{amssymb,tikz}
\usepackage{graphicx}
\usepackage{adjustbox}
\usepackage{xcolor}

\usetikzlibrary{positioning}
  
\tikzset{sgplattice/.style={inner sep=1pt,norm/.style={red!50!blue},char/.style={blue!50!black},
  lin/.style={black!50}},cnj/.style={black!50,yshift=-2.5pt,left=-1pt of #1,scale=0.5,fill=white}}
\usetikzlibrary{shapes.geometric, arrows}

\usetikzlibrary{decorations.markings}
\tikzstyle{startstop} = [rectangle, rounded corners, 
minimum width=3cm, 
minimum height=1cm,
text centered, 
draw=black, 
fill=red!30]

\tikzstyle{io} = [trapezium, 
trapezium stretches=true, 
trapezium left angle=70, 
trapezium right angle=110, 
minimum width=3cm, 
minimum height=1cm, text centered, 
draw=black, fill=blue!30]

\tikzstyle{process} = [rectangle, 
minimum width=3cm, 
minimum height=1cm, 
text centered, 
text width=3cm, 
draw=black, 
fill=orange!30]

\tikzstyle{decision} = [diamond, 
minimum width=3cm, 
minimum height=1cm, 
text centered, 
draw=black, 
fill=green!30]
\tikzstyle{arrow} = [thick,->,>=stealth]

\tikzset{
  symbol/.style={
    draw=none,
    every to/.append style={
      edge node={node [sloped, allow upside down, auto=false]{$#1$}}}
  }
}

\newcommand{\function}[5]{%
  \begin{tikzcd}[
    column sep=2em,
    row sep=1ex,
    ampersand replacement=\&
  ]
  #1\colon \&[-3em]
  #2\vphantom{#3} \arrow[r] \&
  #3\vphantom{#2} \\
  \&
  #4\vphantom{#5} \arrow[u,symbol=\in] \arrow[r,mapsto] \&
  #5\vphantom{#4} \arrow[u,symbol=\in]
  \end{tikzcd}%
}

\def\W{\mathcal{W}^{D+1}}

\def\H{\text{H}}
\def\PD{\mathcal{PD}}
\def\R{\mathbb{R}}
\def\L{\mathcal{L}}
\def\Z{\mathbb{Z}}
\def\F{\mathbb{F}}
\def\e{\text{exp}}
\def\p{{(p)}}
\def\spin{\text{spin}^c}
\def\bbox{\begin{tcolorbox}}
\def\ebox{\end{tcolorbox}}

\def\O{\hat{\cal O}}
\def\Oo{\hat{\cal O}_{\gamma^1}}

\def\be{\begin{equation}}
\def\ee{\end{equation}}
\def\ba{\begin{eqnarray}}
\def\ea{\end{eqnarray}}

\newcounter{mybox}[section]

\begin{document}

\title{A mini-course on Generalized Symmetries and a relation to Cobordisms and K-theory}

\author[]{Oscar Loaiza-Brito} 
\author[]{and V\'ictor M. L\'opez-Ramos}

\affiliation[]{Departamento de F\'isica, Universidad de Guanajuato, Loma del Bosque No. 103 Col. Lomas del Campestre C.P 37150 Leon, Guanajuato, Mexico.}


\date{\today}

\abstract{This mini-course, conducted at the XI School on Geometric, Algebraic, and Topological Methods in Quantum Field Theory held in Villa de Leyva, Colombia, provides an overview of the interconnection between generalized symmetries and cohomology. It is designed for advanced undergraduate students with a background in physics or mathematics. Additionally, we describe a method for connecting generalized symmetries with cobordisms and K-theory within the framework of string theory. We assume basic knowledge in quantum field theory and differential forms.}
\arxivnumber{}


\maketitle

%
%

\section{Introduction}

These lectures constitute an expanded version of a mini-course presented at the XI School on Topological and Algebraic Methods in Quantum Field Theory, held in Villa de Leyva, Colombia. In these lectures, we offer a general perspective on one of the most interesting topics in theoretical physics over the last decade: the study of generalized symmetries. However, it is important to note that these notes do not provide an overall review of this topic. Instead, we present a basic introduction based on fundamental knowledge, assuming that interested students possess a foundational understanding regardless of their background in physics or mathematics. Given the purpose of the School, our aim is to provide a bridge between mathematics and physics, enabling students to delve deeper into the study of symmetries in physics. We also emphasize that our motivation and the structure of our presentation are guided mostly by physics-based arguments. Consequently, we do not offer a formal mathematical construction but instead provide a physically motivated approach to interconnecting ideas and concepts. For this, we assume some basic notions of Quantum Field Theory without overly focusing on precise definitions and notions.\\

The initial part of the notes is dedicated to covering the construction of generalized symmetries from a physical perspective, utilizing well-defined mathematical objects such as differential forms. An important aspect of this discussion is the identification of broken and gauged symmetries with zero cohomology classes. This serves as a central point of discussion as we aim to relate these arguments to recent ideas in quantum gravity and string theory. However, we do not cover the existence of non-invertible symmetries. The second part of the lectures focuses on the use of a particular generalization of de Rham cohomology and its implications for the generalized symmetries defined earlier. In the third part, we discuss and describe various scenarios of importance in string theory by refining the notion of cohomology through the application of the Atiyah-Hirzebruch Spectral Sequence, which maps cohomology to K-theory. The latter provides a mathematical structure that is very useful in the classification of objects in string theory. Based on these scenarios, we present a generalization of the notion of symmetry. Lastly, in the final section, we briefly summarize, based on the results presented throughout the lectures, one of the most recent proposals in the study of quantum gravity within the string theory perspective, known as the {\it cobordism conjecture}.\\

We hope that these notes serve as a starting point for students interested in these topics, whether from a mathematical or physical perspective. References are provided for readers interested in exploring further applications in both physics and mathematics. Additionally, we have included several boxes containing concrete examples, which readers can safely skip if desired.\\

\section{Symmetries in Quantum Field Theory}
In this section, we briefly review some important aspects of the concept of symmetries in classical and quantum  field theory. The references covering this material are quite extensive, so the list we present here is limited. Therefore, we refer the reader to \cite{weinberg1995quantum, ryder1996quantum, zee2010quantum, peskin2018introduction, burgess2020introduction}. The presentation is based on a physical perspective, which may be obscure for a student with a formal mathematical background.\\

\subsection{Field formulation of symmetries}
Let us start by reviewing some standard notation in physics.  As usual, we shall denote the space-time coordinates by $x^\mu$, with $\mu=0,1,\cdots,D$, where the local coordinate $x^0$ denotes the temporal coordinate and $\W$ is a smooth, orientable manifold with Euclidean signature representing a $(D+1)$-dimensional space-time. At a given time $t_i$,  the $D$-dimensional space foliation $\Sigma^D_i$ is also assumed to be {\bf a smooth Riemannian oriented manifold}. As we shall see, under certain circumstances, this space is considered compact or non-compact.\\

At the classical level, a symmetry is described by {\it Noether's theorem} as follows. The presence of symmetries is expressed in terms of a functional called {\it the action} $S$, which in turn is written in terms of another functional, {\it the Lagrangian density} $\L$, 
 depending on the scalar fields $\phi(x^\mu)$ and their derivatives $\partial_\nu\phi(x^\nu)$. For simplicity, we are considering only a set of scalar fields, with $a=1,\cdots, N$. The typical action is then given by
 \be
 S[\phi]=\int d^{D+1}x ~\L (\phi, \partial_\mu\phi),
 \ee
 where 
 \be
 \L=\partial_\mu\phi~\partial^\mu\phi +V(\phi), \\
 \label{Lscalar}
 \ee
 where a summation on the index $\mu$ is implicit and $V(\phi)$ is the scalar potential.\\
 
 Consider now a {\it continuous global transformation} $U \in G$ where $G$ is taken to be a compact continuous group. For the sake of simplicity, we are going to consider this group to be the abelian  Lie group $U(1)$. Hence, the transformation is of the form $U=\text{exp}(i\omega^aT_a)$ where $T_a$ is the generator of the transformation  satisfying the corresponding Lie algebra and $\omega^a$ is a continuous parameter. Therefore,   the transformation on the fields is given by
 \be
 \phi(x^\mu)\mapsto \phi'=e^{i\omega_aT^a}\phi(x).
 \label{phitrans}
 \ee
 By denoting $\delta \phi=i \omega_aT^a(\phi)$, the infinitesimal $U(1)$ transformation acts on the field as
 \be
 \phi \mapsto \phi'= \phi + \delta\phi,
 \ee
 from which one also obtains corresponding variations on $\mathcal{L}$ and $S$ as
 \begin{eqnarray}
 \mathcal{L}(\phi)\mapsto \mathcal{L}(\phi')&=&\mathcal{L}(\phi)+\delta\mathcal{L},\\
 S[\phi]\mapsto S[\phi'] &=&S[\phi]+\delta S.
 \end{eqnarray}
 
The $U(1)$ transformation, being a symmetry, implies that $\delta\mathcal{L}$ can be written as the divergence of a vector $K^\mu$. This is
 \be
 \delta\L(\phi,\partial_\mu\phi)=\partial_\mu K^\mu.
 \ee

By applying the transformation directly to $\L$, a more explicit expression for $\delta S$, up to the Euler-Lagrange equations, is given by
 \be
 \delta S= \int d^{D+1}x~\partial_\mu\left(\frac{\partial\L}{\partial(\partial_\mu\phi)}\delta\phi -K^\mu\right).
 \label{vS}
 \ee
 
 Therefore, there is a {\it continuous global symmetry} if $\delta S=0$  for any infinitesimal parameter $\omega_a$, implying that 
 \be
 \partial_\mu \left(\frac{\partial\L}{\partial(\partial_\mu\phi)}\delta\phi -K^\mu\right)=0.
 \ee
Then it is said that  
\be
J^\mu=\left(\frac{\partial\L}{\partial(\partial_\mu\phi)}\delta\phi -K^\mu\right).
\ee
is a {\it conserved current}. This is one of the most important results in theoretical physics: the {\it  Emmy Noether's theorem}. \\

From (\ref{phitrans}) the current is expressed as $J^\mu=J^{\mu\,a}\,\omega_a$. This allows us to define an important physical quantity that is preserved over time, called a global charge. The global charge is 
\be
Q=\int J^0 d^D x,
\ee
where the temporal component of the current $J^0$ is taken at some specific time for the local observer. In our case, since we are specifying the non-compact space  $\Sigma^D$ according to the times $t_i$ or $t_f$ we shall denote the corresponding conserved charges as $Q_i$ and $Q_f$. It is straightforward to show that 
\be
\frac{\partial Q}{\partial t}=0,
\ee
 for which $Q_i=Q_f$. \\

 Notice that the integration occurs over the spacetime $\W$. For the purposes of these notes, we shall restrict our description to a very specific form of $\W$. We are assuming that at some initial time $t_1$ we perform a measurement of a physical quantity, and we are going to measure it again at some later time $t_f$. If this quantity remains the same, we can say that it stays invariant over time and is conserved. In this context, the spacetime $\W$ is considered a compact portion of the whole spacetime. The boundary of $\W$ is then given by the disjoint union of two space manifolds $\Sigma^D_i$ and $\Sigma^D_f$. It is over these space manifolds that we are integrating in expression \ref{vS}. \\

\bbox[title=Global Symmetry breaking: Goldstone boson, breakable]
Let us present an example of the breakdown of a global continous symmetry. Consider the functional form for a complex scalar field's Lagrangian, denoted as
\begin{equation}\label{11.9}
    \mathcal{L} = \partial_\mu \phi^\ast \partial^\mu \phi - V(|\phi|),
\end{equation}
where the potential \(V\) is defined by
\begin{equation}
    V(|\phi|) = \frac{1}{2} \lambda^2\left(|\phi|^2 - \eta^2\right)^2,
\end{equation}
with $\lambda$ and $\eta$ being real numbers. This potential is known as a double-well potential for the magnitude of \(\phi\), possessing a continuous array of minimum points,
described by \(\langle|\phi|\rangle = \eta\).
The Lagrangian exhibits a global \(U(1)\) symmetry represented by the transformation
\begin{equation}
    \phi \rightarrow e^{i \theta} \phi,
\end{equation}
where \(\theta\) is an arbitrary constant. The scalar field will settle into one of these ground states
leading to the spontaneous breaking of the \(U(1)\) symmetry. 
To decode the theory's spectrum after the breakdown, investigation of minor oscillations around the ground state is essential. Thus, at the minimum, field fluctuations are described by a complex field of the form
\begin{equation}\label{3.41}
    \phi(x) = \eta + \frac{1}{\sqrt{2}}(\chi(x) + i \psi(x)),
\end{equation}
with \(\chi\) and \(\psi\) being real fields.
Notably, the set of ground states forms a circle with radius \(\eta\) in the plane of the complex field. Considering the expansion about the point \((\operatorname{Re} \phi = \eta, \operatorname{Im} \phi = 0)\), \(\chi\) represents a fluctuation perpendicular to the $\psi$-direction, whereas \(\psi\) indicates a tangential fluctuation, as depicted in Figure \ref{higgs}. \\

\begin{center}
\begin{minipage}{0.7\textwidth} 
        \includegraphics[width=0.7\textwidth]{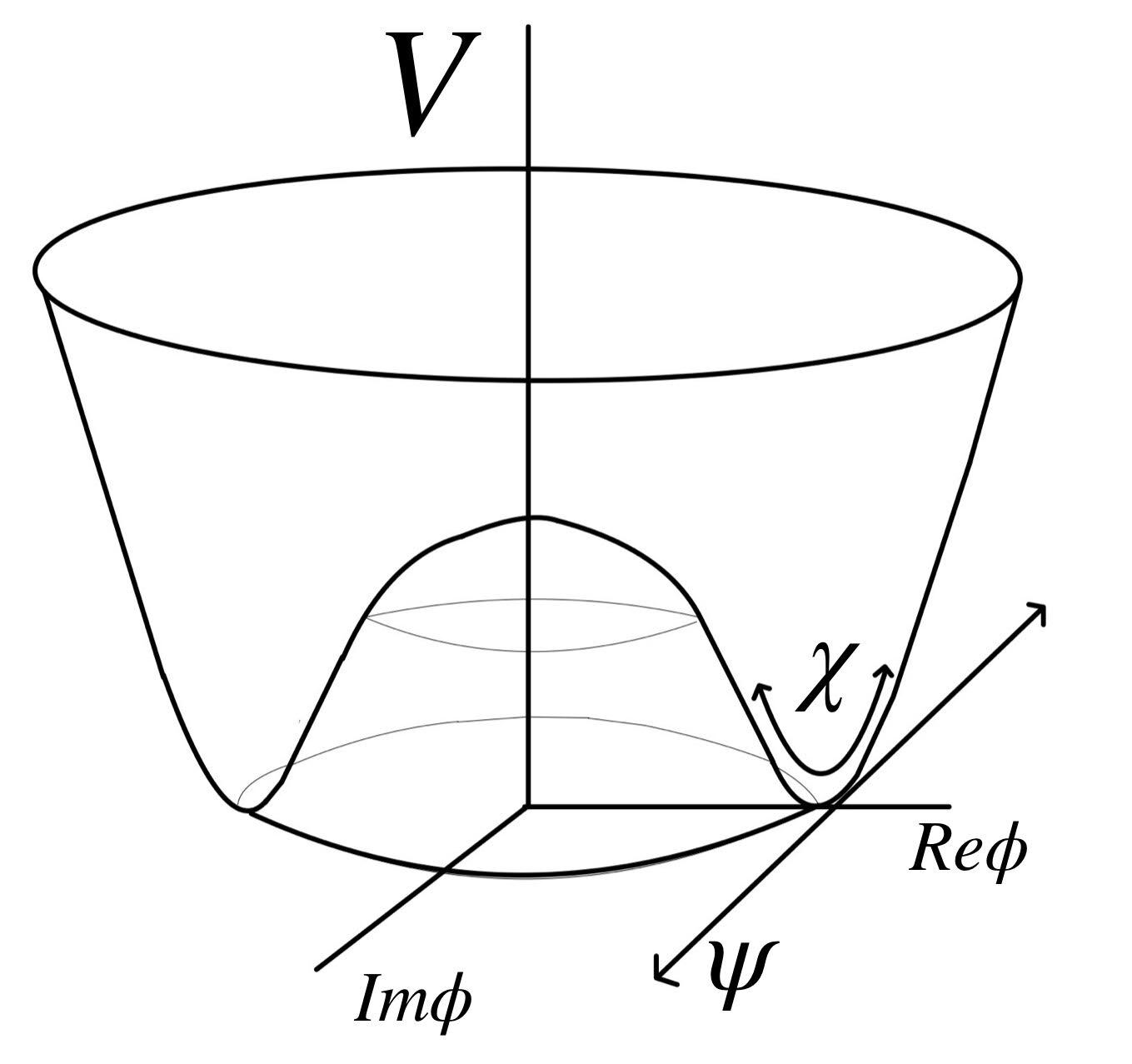} 
        \captionof{figure}{Spontaneous Symmetry breaking}
        \label{higgs}
    \end{minipage}
\end{center}

Around this ground state, the Lagrangian \ref{11.9} becomes
\begin{equation}
\L = \frac{1}{2} \partial_\mu \chi \partial^\mu \chi + \frac{1}{2} \partial_\mu \psi \partial^\mu \psi - \frac{\lambda^2}{8}\left[(2 \sqrt{2} \eta) \chi + \chi^2 + \psi^2\right]^2,
\end{equation}
from which we observe a mass term for \(\chi\),
\begin{equation}
    \frac{1}{2} m_\chi^2 = \frac{\lambda^2}{8}(2 \sqrt{2} \eta)^2 = \lambda^2 \eta^2,
\end{equation}
but none for \(\psi\), indicating that \(\psi\) is a massless field. \\

In summary, within this framework, the spontaneous breaking of $U(1)$ symmetry coincides with the emergence of a massless spin-0 boson in the spectrum, exemplifying the {\it Goldstone theorem}. This theorem posits that in a Lorentz-invariant, local field theory with a Hilbert space having a positive definite scalar product, the spontaneous breaking of a continuous global symmetry results in the appearance of a massless particle, known as a Goldstone (or Nambu-Goldstone) particle, for each symmetry-breaking generator in the expansion around the symmetry-breaking vacuum.\\


\ebox

\subsection{Quantum field formulation of symmetries}
 
The previous discussion on global symmetries can be extended to the quantum level, where Ward identities play a principal role. To introduce them, it is essential to resort to the path integral formulation. Recall that the partition function and correlation functions are expressed, respectively, as
\begin{align}
    Z &= \int \mathcal{D}\phi \, e^{iS}, \nonumber\\
    \langle X \rangle &= \frac{1}{Z} \int \mathcal{D}\phi\, X \,e^{iS},
\end{align}
where $S=S(\phi)$ represents the action of our theory, $X$ a product of fields $\phi$, and $\mathcal{D}$ the functional integration measure. $\phi$ is considered an operator on a Fock space.
Here, $Z$ sums over all possible field configurations and normalizes the
expectation value $\langle X\rangle$. For example, if
$X=\phi(x)\phi(y)$, then
\begin{equation}
    \langle\phi(x)\phi(y)\rangle
    =
    \frac{1}{Z}
    \int\mathcal{D}\phi\,
    \phi(x)\phi(y)e^{iS[\phi]},
\end{equation}
which, for a free scalar field, is the two-point propagator describing the
correlation between the field at $x$ and at $y$.\\

In the path integral formalism, a global continuous symmetry 
given by a transformation on the fields, as in expression (\ref{phitrans}), is reflected in the partition function as
\begin{align}
    Z\rightarrow Z' &= \int \mathcal{D}\phi'\, e^{-S[\phi']}\nonumber\\
    &= \int \mathcal{D}\phi\, e^{-S[\phi]}\left(1+\int d^{D+1}x\,J^{\mu\,a}\,\partial_\mu\omega_a\right).
\end{align}

Being a symmetry, the $U(1)$ transformation on the quantum fields implies that
\begin{equation}\label{wardd}
    \langle \partial_\mu J^\mu \rangle = 0,
\end{equation}
which is the desired current conservation equation.\\

Let us consider a more general operator called {\it a local operator}. A local operator in quantum field theory is an operator constructed from the fields and their derivatives evaluated at a single spacetime point $x$ and is denoted as 
\begin{equation}
    \mathcal{O} (x)\equiv \mathcal{O}(x)\left(\phi_j(x),\partial_{\mu}\phi_j(x),...\right).
\end{equation}

Under an infinitesimal  continuous transformation, a local operator transforms as
\begin{equation}
\mathcal O(x)\longrightarrow
\mathcal O(x)+\delta\mathcal O(x).
\end{equation}
For a symmetry of the action, Noether's theorem gives a conserved current $J^\mu$ at the classical level. Now consider a correlation function of local operators,
\begin{equation}
\left\langle
\mathcal O_1(x_1)\cdots\mathcal O_n(x_n)
\right\rangle
=
\frac{1}{Z}
\int \mathcal D\phi\,
e^{iS[\phi]}
\mathcal O_1(x_1)\cdots\mathcal O_n(x_n).
\end{equation}

Under the field transformation (\ref{phitrans}), the action varies as
\begin{equation}
\delta S
=
\int d^{D+1} x\,
J^{\mu\,a}(x)\partial_\mu\omega_a .
\end{equation}
up to conventions and boundary terms. Since a change of integration variables cannot change the path integral, assuming that the functional measure is invariant, we have
\begin{equation}
0=
\delta
\left\langle
\mathcal O_1(x_1)\cdots\mathcal O_n(x_n)
\right\rangle.
\end{equation}

There are two contributions: one from the variation of the action and another from the variation of the inserted operators. Thus,
\begin{equation}
0=
i\left\langle
\delta S\,
\mathcal O_1\cdots\mathcal O_n
\right\rangle
+
\sum_{k=1}^n
\left\langle
\mathcal O_1\cdots
\delta\mathcal O_k(x_k)
\cdots\mathcal O_n
\right\rangle.
\end{equation}

After integrating by parts and using the arbitrariness of $\omega_a(x)$, one obtains the \textbf{local Ward identity}
\begin{equation}
\partial_\mu
\left\langle
J^\mu(x)
\mathcal O_1(x_1)\cdots\mathcal O_n(x_n)
\right\rangle
=
-i
\sum_{k=1}^{n}
\delta^{(D+1)}(x-x_k)
\left\langle
\mathcal O_1\cdots
\delta\mathcal O_k(x_k)
\cdots\mathcal O_n
\right\rangle ,
\end{equation}
where the precise factor of $i$ depends on the Minkowski or Euclidean conventions. The important point is that the current is conserved away from the operator insertions,
\begin{equation}
\partial_\mu
\left\langle
J^\mu(x)\mathcal O_1\cdots\mathcal O_n
\right\rangle
=0,
\qquad x\neq x_i,
\end{equation}
but when $x$ reaches an insertion $x_i$, a contact term appears. These contact terms encode how each local operator transforms under the symmetry. In this sense, a Ward identity is the quantum-field-theory statement that a symmetry constrains correlation functions.\\

The Ward identity also implies a profound relation between the conserved charge operator $Q$ and the $U(1)$-symmetry.
For that, consider just one operator $\mathcal O(y)$. Then
\begin{equation}
\partial_\mu
\left\langle
J^\mu(x)\mathcal O(y)
\right\rangle
=
-i\,\delta^{(D+1)}(x-y)
\left\langle
\delta\mathcal O(y)
\right\rangle.
\end{equation}

Now integrate this equation over a small spacetime region surrounding the point $y$. By Gauss' theorem,
\begin{equation}
\int_{\Sigma^{D+1}} d^{D+1}x\,
\partial_\mu
\left\langle J^\mu(x)\mathcal O(y)\right\rangle
=
\int_{\partial \Sigma^{D+1}}dS_\mu
\left\langle J^\mu(x)\mathcal O(y)\right\rangle .
\end{equation}

Choose the region to be a thin time slab from $t=y^0-\omega$ to $t=y^0+\omega$. In the limit $\omega\to0$, the spatial boundary does not contribute, and we obtain
\begin{equation}
\left\langle
Q(y^0+\omega)\mathcal O(y)
\right\rangle
-
\left\langle
Q(y^0-\omega)\mathcal O(y)
\right\rangle
=
-i\left\langle
\delta\mathcal O(y)
\right\rangle .
\end{equation}

Because of time ordering, the charge is on different sides of the operator in the two terms,
\begin{equation}
\left\langle
Q\,\mathcal O(y)
\right\rangle
-
\left\langle
\mathcal O(y)\,Q
\right\rangle
=
-i\left\langle
\delta\mathcal O(y)
\right\rangle .
\end{equation}

Therefore,
\begin{equation}
\left\langle
[Q,\mathcal O(y)]
\right\rangle
=
-i\left\langle
\delta\mathcal O(y)
\right\rangle .
\end{equation}

Since this relation holds within arbitrary correlation functions, we identify the operator relation
\begin{equation}
[Q,\mathcal O(x)]
=
-i\,\delta\mathcal O(x)
\end{equation}
or equivalently,
\begin{equation}
\delta\mathcal O(x)
=
i[Q,\mathcal O(x)]
\end{equation}
depending on the convention for the infinitesimal transformation. This is precisely the statement that the conserved charge $Q$ is the generator of the symmetry. Indeed, the corresponding finite transformation is
\begin{equation}
U(\omega)=e^{i\omega Q},
\end{equation}
and an operator transforms as
\begin{equation}
\mathcal O
\longrightarrow
U(\omega)\mathcal O U^{-1}(\omega).
\end{equation}

Expanding to first order,
\begin{equation}
U(\omega)\mathcal O U^{-1}(\omega)
=
\mathcal O
+i\omega[Q,\mathcal O]
+\mathcal O(\omega^2),
\end{equation}
so
\begin{equation}
\delta\mathcal O
=
i[Q,\mathcal O].
\end{equation}

The Ward identity is, therefore, the bridge between Noether's conserved current and the statement that the corresponding charge acts as the generator of the symmetry.\\


\section{Symmetries in Quantum Field Theory and cohomology}

This section is devoted to studying symmetries and conserved charges from the point of view of cohomology.  Some basics on the formalism of differential forms are given in Appendix \ref{apendice}. The following references are highly recommended: 
for some formal mathematical description, see \cite{bott1982differential, hatcher2005algebraic, lawson2016spin, guillemin2019differential}, and for some physically motivated literature, 
see \cite{nash1988topology, nash1991differential, naber1997topology, bertlmann2000anomalies, von2009differential, nakahara2018geometry}.

\subsection{Symmetries as differential forms: 0-form symmetries}

There are a variety of reasons to study symmetries as equivalence classes in cohomology.  Although the correct expressions in QFT are those involving expectation values, we shall omit such notation for simplicity. The reader must keep in mind that we are discussing quantum operators.\\

In QFT, every Lagrangian density $\mathcal{L}$  must be Lorentz invariant (which is usually described in physics as having a term with {\it all indices contracted}, as in the scalar field case shown in Eq.(\ref{Lscalar})). Such an expression is obtained as the field coefficient in the $(D+1)$-form $d\phi\wedge\ast d\phi$, for which the Lagrangian density can be easily expressed as a  closed $(D+1)$-form in the cohomology group $\H^{D+1}(\W)$, i.e.,
\be
\L_{D+1}=d\phi\wedge\ast d\phi  \,\in \,\H^{D+1}(\W),
\ee
where the action is given by
\be
S=\int_{\W}\L_{D+1},
\ee
and we are considering $p$-differential forms composed of quantum states, which in turn take values in the Hilbert space ${\cal H}$. Under a transformation of the fields $\phi$ or the local coordinates $x^\mu$, the action as well as $\L$ can suffer a variation.\\

Then there is  a {\it symmetry} if the Lagrangian varies as
\be
\L\longrightarrow \L +d\ast L_1,
\ee
where $d\ast L_1$ is actually $\delta\L$. 

\begin{tcolorbox}[title=Symmetry, label=symmetry]
A continuous global symmetry is a transformation $\hat{U}$ acting on the quantum fields $\phi \in {\cal F}$, where ${F}$ is the Fock space, generated by elements of a continuous compact Lie group (think here as $U(1)$) such that if $\L_{D+1}$ defines a class in $\H^{D+1}(\W;\Z)$, the transformed Lagrangian $\L'=\L(\phi', d\phi')$ still belongs to the same equivalence class $[\L_{D+1}]\in \H^{D+1}(\W;\Z)$. Therefore the variation of the Lagrangian closed form is an exact form.\\

{\it Why does the Lagrangian and its variations define an integral cohomology class instead of a real one?} Observe that for a generic symmetry, the associated charge may take any real value. If, however, we require the global structure of the symmetry group to be compact, as in the case of a $U(1)$ symmetry, then, with $Q$ denoting the generator of the transformation, its possible values become quantized to the integers. To illustrate this, let us look at the simplest situation involving a single generator. In this case, the transformation takes the form
\[
\hat{U}(\alpha)=e^{i\alpha \hat{Q}}.
\]
Since $\hat{U}(\alpha+2\pi)=\hat{U}(\alpha)$ we conclude that $Q\in\Z$. Notice that the integer quantization comes from the fact that $U(1)$ is compact. Mathematically this means that 
\[
\frac{1}{2\pi}\int_{\Sigma^D}\ast J_1\in \Z
\]
for every closed $D$-cycle $\Sigma^D$. Therefore the equivalence class $[\frac{\ast J_1}{2\pi}]$ is not merely a class in the de Rham cohomology $\H^D_{dR}(\W)$ but comes from an integral cohomology class
\[
\H^D(\W;\Z)\rightarrow \H^D(\W;\R)\simeq H_{dR}^D(\W).
\]
\end{tcolorbox}
\vspace{0.5cm}

Under the operator $\hat{U}$,  the action transforms as
\be
S\longrightarrow S+\delta S,
\ee
where $\delta S=0$ if the transformation is a global symmetry. This means that we can write
\be
\delta S=\int_{\W}d\ast K_1,
\ee
or by actually transforming the Lagrangian form $\L_{D+1}$ as
\be
\delta S=\int_{\W}d\ast L_1,
\ee
where $[d\ast K_1]=[0]\in \H^{D+1}(\W)$.  This defines the closed $D$-form Noether current $\ast J_1$ as
\be
\ast J_1=\ast L_1-\ast K_1.
\ee
Notice that this is physically interpreted as having a conserved current,
\be
\partial_\mu J^\mu=0 \qquad \leftrightarrow  \qquad d\ast J_1=0,
\ee
where $\ast J_1=\frac{1}{6}J_\mu \,\epsilon^\mu_{\,\,\,\alpha\beta\gamma}dx^\alpha \wedge dx^\beta\wedge dx^\gamma$.\\

%
%
 Expressing the conservation of a physical quantity in this way leads us to a beautiful way to interconnect physics and cohomology. The current conservation implies that the current $\ast J_1$ is a closed  $D$-form and therefore belongs to an equivalence class in the corresponding cohomology group; this is
 \be
 [\ast J_1]\in H^D(\W;\F).
 \ee
 The current class $[\ast J_1]$ can be a zero or a non-zero element in $\H^D(\W;\F)$, and this is reflected in the number associated with this class through the isomorphism \ref{iso}, from which the conserved charge is defined as
 \be
 Q_i=\int_{\Sigma^D_i}\ast J_1,
 \ee
where $\Sigma^D_i$ is the space slice taken at some time $t=t_i$.\\

Therefore, the conserved charge $Q_i\in \text{Hom}(\H^D(\W;\F), \H_D(\W;\F))$ is indeed a topological invariant that depends uniquely on the equivalence classes.  Physically, we can enumerate some consequences and implications that follow:
 \begin{enumerate}
 \item The global charge $Q_i$ requires measuring the current $\ast J_1$ throughout the space $\Sigma^D_i$ (this is quite different from a local gauge charge, as we shall  shortly see).
 \item At some point in time $t_i$, we measure the charge $Q_i$ through the space $\Sigma^D_i$. After some time $\Delta t=t_f-t_i$, we again measure the charge through a homotopically deformed $\Sigma^D_f$. Since $\Sigma^D_i$ and $\Sigma^D_f$ both belong to the same equivalence class in $\H_D(\W;\Z)$, the charge is conserved. See Figure \ref{Fig1}, where $\partial\W=\Sigma^D=\Sigma^D_i\sqcup \overline{\Sigma}^D_f$ and the bar on $\overline{\Sigma}^D_f$ refer to the space $\Sigma^D_f$ with the opposite orientation.
  \begin{figure}[h]
  \centering
  \includegraphics[width=0.5\textwidth]{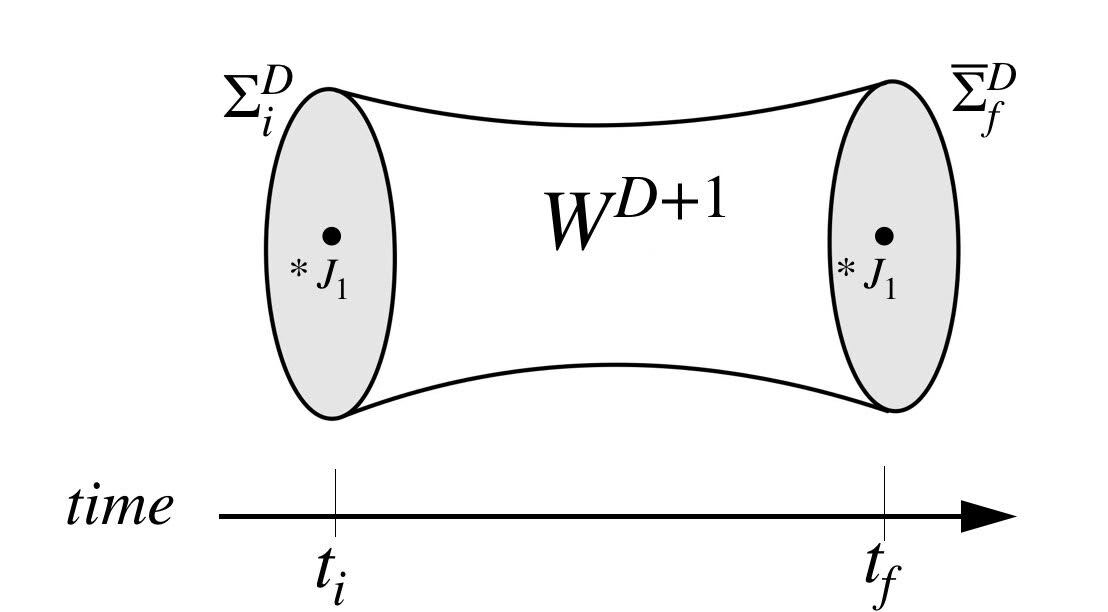}
  \caption{Conservation of a global charge.}
  \label{Fig1}
\end{figure}
  \item We can also consider $\Sigma^D_i=\sqcup_{j=1}^M \Sigma^D_{ij}$, which consists of measuring $M$ initial global charges. Having a symmetry means that the charge $Q_i=\sum_j Q_{ij}$ is preserved  when measuring at time $t_f$. For simplicity, we shall limit our study to $M=2$.
 \item We define the {\it total global charge} $Q_T$ by
 \be
 Q_T=\int_{\W}d\ast J_1=\int_{\partial\W}\ast J_1.
 \ee
 Having a global  symmetry implies that $Q_T=0$. Since $\Sigma^D=\Sigma^D_i\sqcup\overline{\Sigma}^D_f$, the global charge is conserved.
 \item
 By Poincar\'e Lemmas, the charge $Q_i$ vanishes if the current $\ast J_1$ is an exact form or if the cycle $\Sigma^D_i$ has a boundary; that is, if we take the isomorphism between $\H^D$ and $\H_D$ over the zero classes. However, this trivial remark establishes some very interesting consequences in physics, as we shall see.
 \item A Symmetry is broken if $d\ast J_1\ne 0$ this is if $\ast J_1$ under a transformation is no longer closed. In this case $\ast J_1 \in C^D$, it does not define a class in $\H^D$ , although it is in the zero class $[0]\in \H^{D+1}$.
   \end{enumerate}
 
 \subsubsection{A conserved global charge}
 Let us, for the moment, concentrate only on a {\bf conserved non-zero global charge}. Usually in physics, we talk about an object carrying the charge, referred to as a particle, or a 0-dimensional object. Let us see how this association is constructed from the cohomology point of view. Given a non-trivial cycle $\Sigma^D_i$ defining an equivalence class in $\H_D(\W;\Z)$ we can use the $\PD$ and Hodge $\ast$-maps to get
 \be
 Q_i=\int_{\Sigma^D_i}\ast J_1=\int_{\W} \ast J_1\wedge \PD (\Sigma^D_i)=\int_{\gamma^1} \PD(\Sigma^D_i) \int_{\Sigma^D_i} \ast J_1,
 \ee
 where we have identified $\gamma^1=\PD(\ast J_1)$ and $\W$ as the space-time generated by the product of $\Sigma^D_i$ and $\gamma_1$. Hence, for each closed form $J_1$, we can associate a 1-dimensional  cycle $\gamma_1$ defining an equivalence class in $\H_1(\W;\Z)$; this is $\partial\gamma^1=0$. This is the world-line of a particle with a current $\ast J_1$. Also,  we can define the {\it intersection number}
 \be
 \text{Int}(\Sigma^D_i, \gamma^1)=\int_{\gamma^1}\PD (\Sigma^D_i)= \int_{\W}\PD (\Sigma^D_i)\wedge \PD(\gamma^1)\in \Z.
 \ee
 For the simplest case, this integral is equal to 1, indicating the presence of a single particle. For the case of $N$ particles, the intersection number would be equal to $N$. In this context, we can say that the symmetry is present since the space $\Sigma^D_i$  does not collapse into a point. Measuring the global charge $Q_i$ at time $t_i$ gives the same result as measuring the global charge $Q_f$ at time $t_f$; this is $[\Sigma^D_i]=[\Sigma^D_f]\ne [0]$.\\

\begin{figure}[h]
  \centering
  \includegraphics[width=0.5\textwidth]{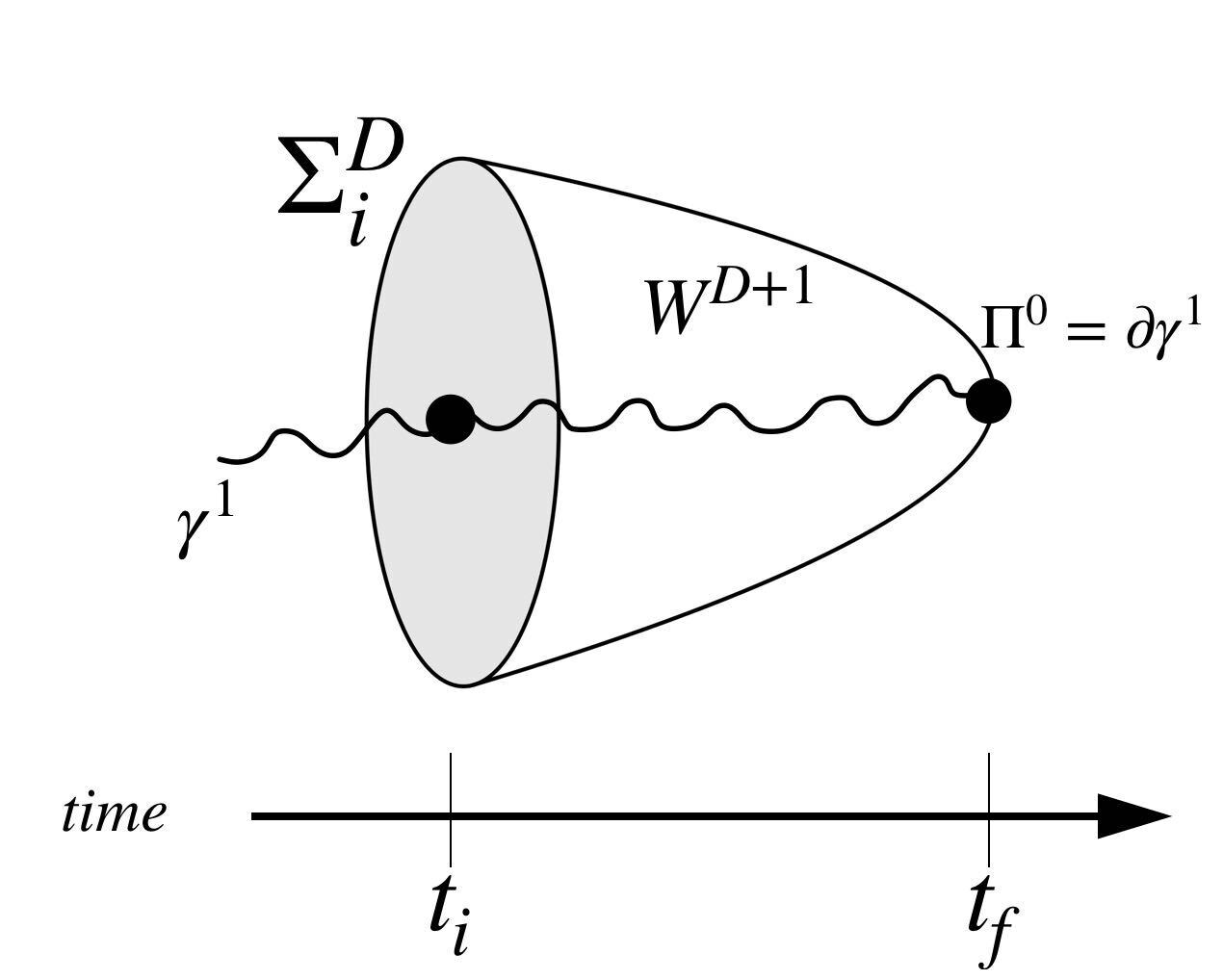}
  \caption{Conservation of a null global charge.}
  \label{Fig2}
\end{figure}

 Consider now the case in which the {\bf global conserved charge $Q_i$ vanishes}.  This happens for $[\Sigma^D_i]=[0]$ or $[\ast J_1]=[0]$. In the first case
 \be
 Q_i=\int_{\Sigma^D_i}\ast J_1=\int_{\W}d\ast J_1=0,
 \ee
 where $\Sigma_i^D$ is the boundary of the space-time $\W$ (contrary to the disjoint union of two or more slices of space). See Figure \ref{Fig2}.\\
 
In the second case, a null global charge $\ast J_1$ is an exact form belonging to the zero class in $\H^{D}(\W;\Z)$. Therefore, a zero global charge is specified by the existence of  a $(D-1)$-form $\ast F_2$ in $C^{D-1}$, such that

 \be
 \begin{tikzcd}
   \cdots \arrow{r}{d} &C^{D-1} \arrow{r}{d}&C^D \arrow{r}{d} & C^{D+1} \arrow{r}{d}& 0 \\
   &\ast F_2 \arrow[r, mapsto]& d\ast F_2=\ast J_1\arrow[r, mapsto]& d\ast J_1=0&&
   \end{tikzcd}
 \ee
 
 The vanishing of the global charge at a given time $t_i$ follows from the exactness of the current $\ast J_1$ and  defines a very important concept in physics known as {\it gauge symmetry} (although, as we shall see, it is not a symmetry in this context). Notice that the equation $\ast J_1=d\ast F_2$ is exactly the same as the inhomogeneous pair of Maxwell equations after appropriate identifications.\\

\bbox[title={Maxwell Equations}, label=maxwell, breakable]
 In terms of vector fields, Maxwell equations (speed of light $c=1$) are commonly written in the physics literature as
 \begin{eqnarray}
\text{inhomogeneous} \qquad \nabla\cdot\vec{E}=\rho, \qquad &&\nabla\times\vec{B}-\frac{\partial\vec{E}}{\partial t}=\vec{J},\\
\text{homogeneous}\qquad \nabla\cdot \vec{B}=0, \qquad && \nabla\times\vec{E}+\frac{\partial\vec{B}}{\partial t}=0,
 \end{eqnarray}
 where the electric and magnetic fields $(\vec{E}, \vec{B})$ are given in terms of the electric and magnetic potentials, $\varphi$ and $\vec{A}$, respectively, as
 \begin{eqnarray}
 \vec{E}&=&-\nabla\varphi-\frac{\partial\vec{A}}{\partial t}, \\
 \vec{B}&=&\nabla\times\vec{A},
 \end{eqnarray}
  with $\rho$ and $\vec{J}$ representing the charge and current densities. In relativistic notation, it is common to express the potentials as a vector in spacetime with coordinates $A^\mu=(\varphi, \vec{A})$, while the derivative operator $\partial_\mu=(\partial_t, \nabla)$.  The same is done for a space-time vector  for the electric charge and current densities, $J^\mu= (\rho,  \vec{J})$. It turns out that $\vec{E}$ and $\vec{B}$ can be arranged in an antisymmetric square matrix  $F$ with components $F_{\mu\nu}$ (known as Faraday's tensor)
  \be
  F_{\mu\nu}=\partial_\mu A_\nu-\partial_\nu A_\mu,
  \ee
  where
  \be
F= \begin{bmatrix}
0& -E_1 & -E_2 &-E_3\\
E_1&0&B_3&-B_2\\
E_2&-B_3&0&B_1\\
E_3&B_2&-B_1&0
\end{bmatrix}.
\ee
The electromagnetic potentials and the space-time current density define 1-forms in $C^1(\W)$ given by
\be
A_1=A_\mu dx^\mu, \qquad J_1=J_\mu dx^\mu,
\ee
while the Faraday's tensor defines a differential 2-form,
\be
F_2=\frac{1}{2}F_{\mu\nu}dx^\mu\wedge dx^\nu.
\ee
It follows that  $F_2=d A_1$, defining a zero class  in $\H^2(\W;\mathbb{R})$. Notice that $F_2 \in C^2$ is well-defined for $A_1\in C^1/(\text{ker}~d)$ (see Section \ref{gaugedsym}). This is known in physics as {\it a gauge symmetry}. Then,  Maxwell's equations can be written in terms of differential forms as
\begin{eqnarray}
\text{inhomogenous:}&\qquad& d\ast F_2=\ast J_1,\\
\text{homogenous:}&\qquad& dF_2=0,
\end{eqnarray}
where $F_2=dA_1$. Notice that in this language, we can say that:
\begin{enumerate}
\item The inhomogeneous Maxwell's equations are described by  a non-closed 2-form $\ast F_2$, or equivalently by a current defining a zero class $[\ast J_1]=[0]\in H^3(\W;\mathbb{R})$.
\item The homogeneous Maxwell's equations are described by the zero class in $\H^2(\W;\mathbb{R})$ defined by $[F_2]$.
\item Charge conservation, described by the equation $d\ast J_1=0$, follows from the fact that the current defines a zero class $[\ast J_1]=[0]\in H^3(\W;\mathbb{R})$.
\item Global charge $Q=\int_{\Sigma^3}\ast J_1$ vanishes since $\ast J_1$ is exact. But {\it electric charge} is not the Global charge. It is then necessary to properly define the concept of electric (and magnetic) charges.
 \end{enumerate}
\ebox

 \subsubsection{A gauged symmetry}\label{gaugedsym}
 A symmetry is gauged when the associated current $\ast J_1$ is exact; that is, when $[\ast J_1]=[0]\in \H^D(\W;\Z)$.  However, the exactness of the current $\ast J_1$ can be seen as an equation of motion derived from a specific Lagrangian given by\footnote{We are ignoring the coupling constants.}
 \be
 \L=F_2\wedge\ast F_2 +A_1\wedge\ast J_1,
 \ee
 where $F_2=dA_1$ is defined to be an exact form, for which $[F_2]=[0]\in\H^2(\W;\Z)$. The implicit equivalence relation defining the equivalence class is known in physics as {\it gauge symmetry}, i.e., a redefinition on $A_1$ which leaves invariant the {\it observable} field $F_2$, with
 \be
 A_1\in C^1(\W;\Z)/\{\text{im } d:C^0\rightarrow C^1\}.
 \ee
 A gauge symmetry is also known as an {\it internal} symmetry, a term which emphasizes that it is not a transformation involving physical coordinates such as space or time.\\
 
By gauging the current, or equivalently making it exact, the global charge at a given time vanishes since, by Poincar\'e's lemmas
   \be
 Q_i=\int_{\Sigma^D_i}\ast J_1=\int_{\Sigma^D_i} d\ast F_2=0.
 \ee
However, we can instead define a gauged charge by only integrating on a compact submanifold $\widetilde{\Sigma}^D_i$ in $\Sigma^D_i$ such that $\partial\widetilde\Sigma_i^D=S_i^{D-1}$. Define {\it the electric (gauged) charge} $q_i^{(e)}$ at time $t_i$ by
 \be
 q^{(e)}_i=\int_{\widetilde\Sigma^D_i}\ast J_1=\int_{S_i^{D-1}} \ast F_2.
 \label{electricq}
 \ee

\begin{figure}[h]
  \centering
  \includegraphics[width=0.5\textwidth]{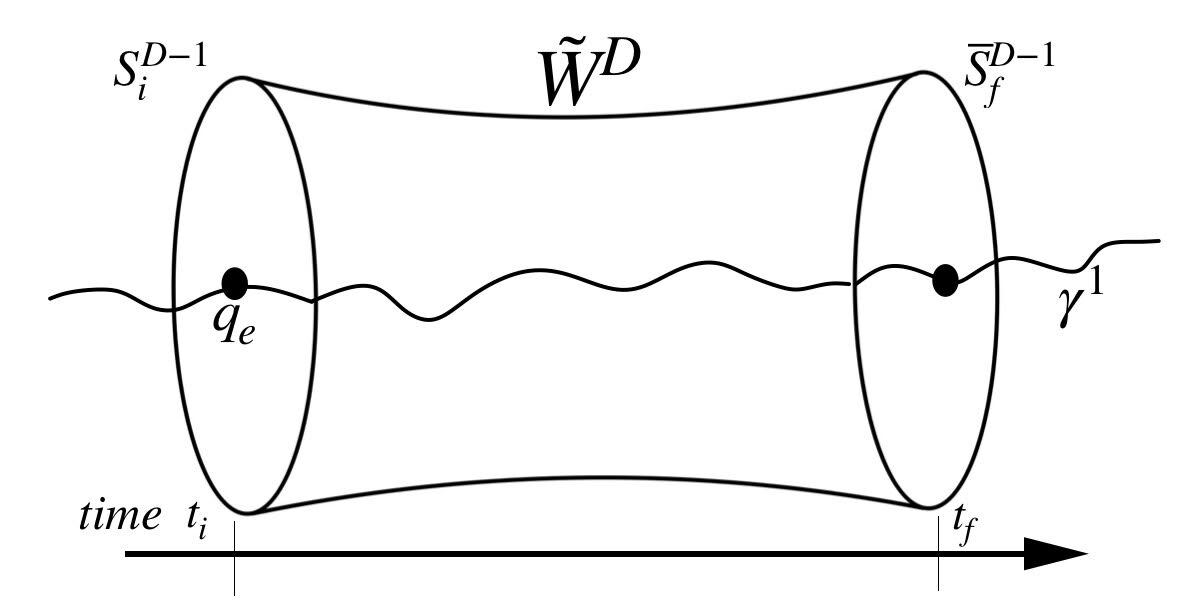}
  \caption{Electric charge conservation.}
  \label{Fig3}
\end{figure}

 In Figure \ref{Fig3}, we represent the conservation of electric charge.   Similarly to the global charge, we can extend the integral to the whole compact space $\widetilde{\Sigma}_i^D$ such that
\be
 q_i^{(e)}=\text{Link}(S^{D-1}_i,\gamma^1)\int_{S^{D-1}_i}\ast F_2,
 \ee
 where $\text{Link}(S^{D-1}_i,\gamma^1)$, the {\it linking number}  between the boundary $S^{D-1}$ and the worldline $\gamma^1$, is defined as
 \be
\text{Link}(S^{D-1}_i, \gamma^1)=\int_{S^{D-1}_i}\PD(\gamma^1),
\ee
is a topological invariant reflecting the number of involved particles carrying the electric charge with $\widetilde{W}^D=S^{D-1}\times \gamma^1$.\\
 
 It is important to notice that the calculation of the electric gauge charge only requires integrating over a closed surface containing the particle, contrary to the global charge case. This is the reason that for a Black Hole with electric gauge charge (Reisner-Nordstr\"om exact solution), we can measure the amount of charge outside the horizon since we only need to measure the flux of the field $\ast F_2$  through the region $S^{D-1}$. However, if a Black Hole or a remnant of it  carries some global charge, we cannot measure it since the space beyond the horizon is not accessible to an external observer. See section \ref{conjecture} for  some relevant uses of these arguments in quantum gravity.\\

 In the same context, we can now define the {\it magnetic charge} as
 \be
 q^{(m)}_i=\int_{S^2}  F_2.
 \ee
 In this case, the magnetic current is $J_2=\ast \widetilde{J}_{D-1}$, defining the corresponding class in $\H^2(\W;\Z)$.\\
 
The existence of monopoles or gauge fields interacting with matter requires extra mathematical structures on the cycles $\Sigma^D_i$. These structures and their corresponding descriptions are extensive, so we shall summarize the main ingredients we need to keep in mind for the generalizations we shall review in the next section.\\

\begin{enumerate}
\item
The spatial cycles $\Sigma^D_i$ must also be base manifolds of a $U(1)$-principal bundle,
where the gauge fields $A_1$ are known as {\it connections}  and $F_2$ as the associated curvature. A trivial bundle (globally the product of the fiber and the base manifold) has zero curvature, implying that $dF_2=0$. The Faraday's 2-form defines a zero class in cohomology, which in turn tells us that the magnetic charge is null.
\item
The charge is defined through a map between the principal bundle $E$ and $\H^\bullet(\W)$. Let us define the group of principal bundles $E$ with base manifold $X$ and fiber $U(1)$ as $K(X)$. This is called the {\it K-theory} group.  Define the {\it Chern character} of the vector bundle $E$, $ch(E)$, as
 \be
 \begin{tikzcd}
ch: K(X) \arrow{r}{}&H^\bullet(\W) \\
   E \arrow[r, mapsto]&ch(E)= \e\left(F_2/2\pi\right)=\sum_i ch_i(E),
   \end{tikzcd}
 \ee
 where $ch_i(E)$ is called {\it the Chern character} with $ch_i(E)\in \H^{i}(\W;\Z)$. Take $X=S^2$, then the gauged  electric charge is given by
 \be
 q_i^{(e)}=\int_{S^{D-1}_i}\ast ch_{1}(E),
 \ee
 where $ch_{1}(E)=[F_2]\in H^{2}(\W;\Z)$. If $[F_2]=0$, there is no magnetic charge since $F_2$ is exact ($dF_2=0$). For a non-trivial class, the magnetic charge is an integer expressed in terms of the Chern character as
 \be
 q^{(m)}_i=\int_{S^2}ch_1(E).
 \ee 
 Then having a magnetic monopole implies that  the principal $U(1)$-bundle over $S^2$ is not trivial (i.e., it is not globally the product of $S^1\times S^2$).
\item
We are considering an orientable base $\Sigma^D_i$. This also implies an extra condition on it. Consider the tangent bundle over $\Sigma^D_i$, $T\Sigma^D_i$, with a Riemannian metric with a group structure $O(D)$. We say that $\Sigma^D_i$ is {\it orientable} if $O(D)$ reduces to $SO(D)$, which occurs when
the {\it first Stiefel-Whitney class} $w_1$, defined as
 $
 \begin{tikzcd}
w_1: K(\Sigma^D_i) \arrow{r}{}&H^1(\Sigma^D_i;\Z_2),
   \end{tikzcd}
$
is trivial.
\item
If the current $\ast J_1$ is composed of fermions (as physically expected), the principal $U(1)$-bundle requires an extra structure called  {\it spin} to properly define them. Basically, a spin structure on $\Sigma^D_i$ implies the possibility to define functions that transform under the double cover of $SO(D)$, which is actually the spin group. Physically, this points out that we are considering fermions. The admittance of spin structures is measured by the {\it second Stiefel-Whitney class} $w_2(\Sigma^D_i)\in \H^2(\Sigma^D_i;\Z_2)$.  If the bundle over $\Sigma^D_i$ accepts a spin structure, we say that it is a spin bundle.  Therefore, there is a spin bundle over $\Sigma^D_i$ if and only if $w_2(\Sigma^D_i)$ is trivial.
\item
The presence of a gauge symmetry requires the interaction between gauge fields $A_1$ and matter through the action term $A_1\wedge\ast J_1$ in the corresponding Lagrangian. Mathematically, this tells us that the principal bundles with base manifold $\Sigma^D_i$ must admit spin structures compatible with the gauge connections $A_1$. This extra structure is called $\spin$. A $\spin$-structure is a pair $(P,\rho)$, where $P$ is a principal spin-bundle over $\Sigma^D_i$ and $\rho$ is a complex representation of the Clifford algebra bundle, associated with $P$ on the complexified tangent bundle of $\Sigma^D_i$. A manifold $\Sigma^D_i$ is said to be $\spin$ if it admits a $\spin$-structure. If this happens, the {\it Third Stiefel-Whitney class}, $w_3(\Sigma^D_i)\in \H^3(\Sigma^D_i;\Z_2)$, is trivial (but not otherwise).\\
\end{enumerate}

 \begin{tcolorbox}[title=Dirac's monopole, label=monopole, breakable]
 The absence of magnetic charge in Maxwell's equation is reflected by the equation $dF_2=0$. This equation follows by considering an exact Faraday's form given by $F_2=dA_1$. Mathematically this means that $A_1$ is globally defined on the whole space-time. The existence of a magnetic charge in Maxwell's equation is then translated into having a non-closed 2-form $F_2$,
 \be
 d F_2= J^{(m)}_3.
 \ee
 However, since Maxwell's theory is actually a gauged theory, it is important to maintain $F_2=dA_1$. This might seem contradictory. The crucial point from a physics perspective is to understand that $A_1$ is not globally defined. Essentially, let us consider the sphere $S^2$ as the combination of two contractible open sets: the upper hemisphere $U$ and the lower hemisphere $V$, which intersect in the equatorial band $U \cap V$. We define a local $A_1^U$ on the upper hemisphere and $A_1^V$ on the lower, such that both differ by a gauge transformation
 \be
 A_1^U-A_1^V=d\Lambda,
 \ee
 where $d\Lambda$ is not exact in $U\cap V$. Hence
 \begin{eqnarray}
 \oint_{S^2}F_2 &=&\int_{S^1} d\Lambda,
 \end{eqnarray}
 and for $\Lambda=q^{(m)}\theta/2\pi$, we get that
 \be
 \oint_{S^2} F_2=q^{(m)}.
 \ee
 The use of the open spaces $U$ and $V$ where we define different gauge fields, can be formally described by the {\it Mayer-Vietoris Sequence}. A magnetic monopole is present if $F_2$ defines  a non-zero equivalence class of $\H^2(S^2;\Z)$. Notice that we are not computing the cohomology group for $\Sigma^3\cong \mathbb{R}^3$, but for $S^2$.  
 For $S^2$, the Mayer-Vietoris sequence for cohomology states 
\[
\cdots \rightarrow H^{n-1}(U \cap V) \rightarrow H^n(S^2) \rightarrow H^n(U) \oplus H^n(V) \rightarrow H^n(U \cap V) \rightarrow H^{n+1}(S^2) \rightarrow \cdots 
\]
which reduces to
\[
0 \rightarrow H^1(S^2) \rightarrow H^1(U) \oplus H^1(V) \rightarrow H^1(U \cap V) \rightarrow H^2(S^2) \rightarrow 0 .
\]
Since $U$ and $V$ are contractible, $H^1(U) = H^1(V) = 0$. Also, $U \cap V$ is homeomorphic to $S^1$, so $H^1(U \cap V)$ is the first cohomology group of the circle, which is isomorphic to $\mathbb{Z}$. Thus, the sequence simplifies to
\[
0 \rightarrow H^1(S^2) \rightarrow 0 \oplus 0 \rightarrow \mathbb{Z} \rightarrow H^2(S^2) \rightarrow 0 .
\]
From this, we can see that $H^1(S^2) = 0$ and $H^2(S^2) \cong \mathbb{Z}$. Therefore, the second cohomology group of the 2-sphere $S^2$ is isomorphic to the integers $\mathbb{Z}$.\\

A more general physical approach for the magnetic monopole is given by the t'Hooft-Polyakov monopole. A pedagogical description is provided in \cite{nash1988topology}.\\
 \end{tcolorbox}

 \subsubsection{Broken Symmetry}
 The last case to analyze is when the symmetry is broken. This translates to the fact that the current $\ast J_1$ is not closed, implying that it cannot define a class in cohomology. Therefore, we can say at most that $\ast J_1\in C^D(\W)$ and 
 \be
[ d\ast J_1]=[0]\in \H^{D+1}(\W;\Z).
\ee
Using the Poincar\'e dual map, a broken symmetry is associated with the existence of a current world line with a boundary; that is,
\be
d\ast J_1= G_{D+1} \mapsto \partial \gamma_1=\pi_0.
\ee
Physically, this is interpreted as a 0-dimensional particle carrying the global charge, with a world line that ends at some time $t$. Since the global charge $Q_i$ disappears (the current is not conserved), the cycle $\Sigma_i^D$ becomes trivial and shrinks into a point. In other words, $\Sigma_i^D=\partial\W$ and $\Sigma_i^D$ are then homologous to a point. See Figure \ref{Fig2}. It is important to notice that a broken symmetry or a conserved null charge both allow the cycle $\Sigma^D$ to be contracted into a point.  \\

\subsection{ Symmetry operators}

At the level of QFT, a global symmetry operator $\hat{U}$ acts on a quantum state in Hilbert space, and it is given by
\be
\hat{U}_g(\Sigma^D_i)=\e (-i\alpha \hat{Q}_i)=\e (-i\alpha\int_{\Sigma^D_i} \ast J_1)
\ee
for $g\in U(1)$, $\alpha$ a continuous parameter and $\hat{Q}$ the generator of the $U(1)$-transformation. For a global symmetry, we extend the current's integral to the whole spacetime by incorporating the intersection number $\text{Int}(\Sigma^D_i, \gamma^1)$, which takes values over the integers,
\be
\hat{U}_g(\Sigma^D_i)=\e\left(-i\alpha \,\, \text{Int}(\Sigma^D_i,\gamma^1)\, \hat{Q}_i\right).
\ee
Notice that a change in the value of the intersection number $\text{Int}(\Sigma^D_i,\gamma^1)$ implies the breakdown of a symmetry.\\

Since the operator $\hat{U}_g$ only depends on the class $[\Sigma^D_i]$, we say that it is a {\it topological operator}. If such a state carries a global charge, we refer to the quantum state as a {\it charged operator} denoted $\O$. If the charge is referred to a gauge symmetry, then the charge operator is called a Wilson Line, as we shall shortly see.\\

\begin{figure}[h]
  \centering
  \includegraphics[width=0.5\textwidth]{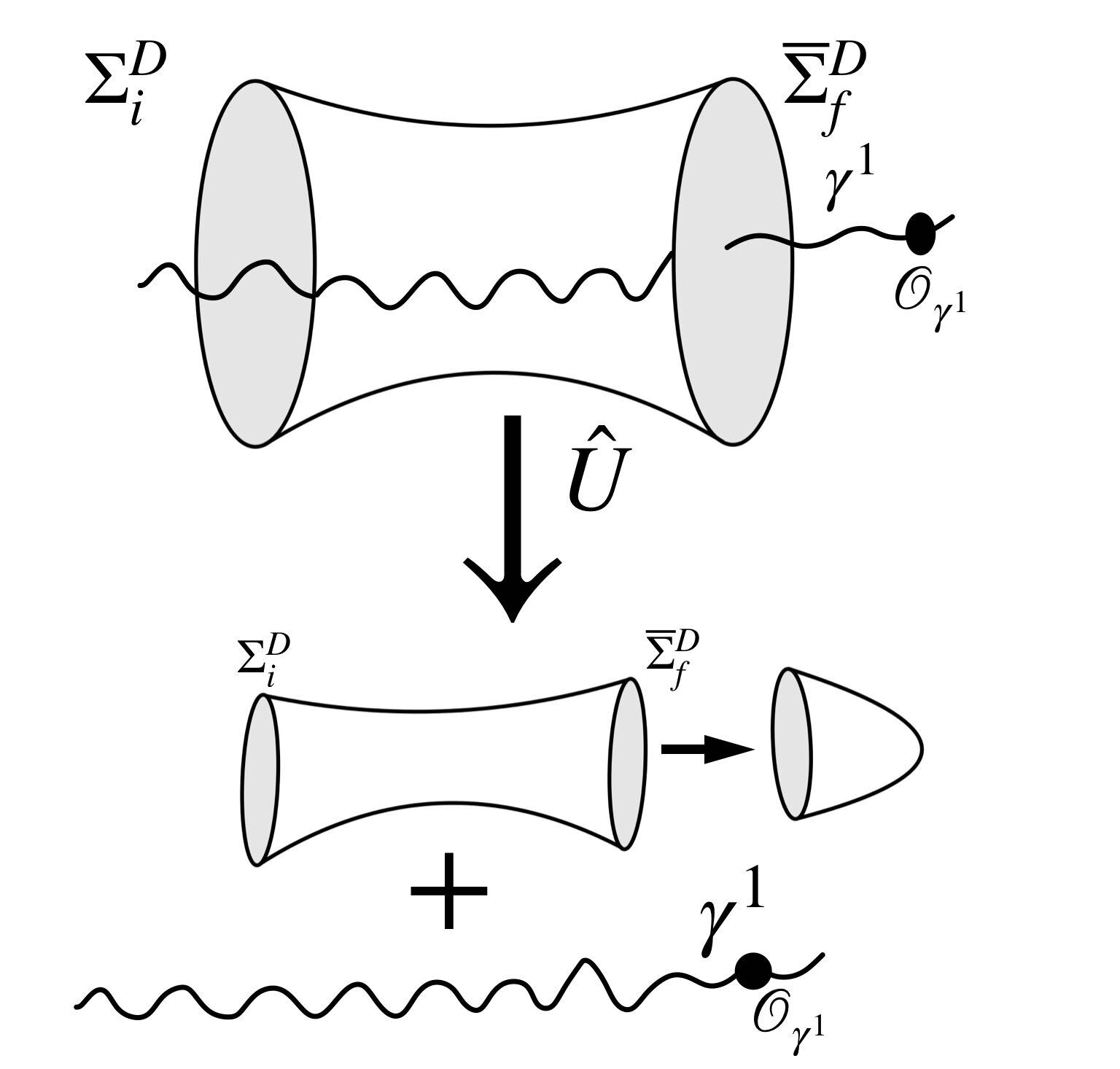}
  \caption{Action of $\hat{U}_g({\Sigma}^D_f)$ on a charge operator.}
  \label{Fig4}
\end{figure}

The charge assigned to an operator state is computed by surrounding the operator with a compact closed submanifold $\widetilde\Sigma^D_i$. The transformed charged operator $\O$ satisfies the relation
\be
\O ''=\hat{U}_g(\Sigma^D)_i\O \hat{U}_g(\Sigma^D_i)^{-1}=R(g)\O,
\ee
where $R$ is an appropriate  group representation. For the case of a continuous compact group, as $U(1)$, we can write
\be
\hat{U}_g(\Sigma^D_i)\O=e^{i\alpha Q}\O.
\ee
Let us now fix some notation. If a quantum state operator ${\cal O}$ corresponds to a particle state describing a world-line $\gamma^1$, we shall denote such a state as $\O_{\gamma^1}$. Consider the case  where the quantum state does not carry a global charge. Then
\be
\hat{U}_g(\Sigma^D_i)=\hat{\text{I}};
\ee
since $\Sigma^D_i$ does not contain a global charge. On the contrary, if the 0-dimensional object (matter) carries some non-zero global charge,
\be
\hat{U}_g(\Sigma^D_i) \Oo=\text{exp}\left(-i\alpha~\text{I}(\gamma^1,\Sigma^D_i)\int_{\Sigma^D_i}\ast J_1\right)\Oo=R(g)~\Oo \hat{U}_g^{-1}(\Sigma^D_i).
\ee
This relation can be interpreted as follows: the symmetry operator $\hat{U}$ transforms the charged operator placed in the interior of $\widetilde\Sigma^D$, which is not contractible precisely because of the presence of the charge operator (i.e., $\text{Int}(\Sigma^{D}_i, \gamma^1)\ne 0$), to a state in Hilbert space that is not in the interior of ${\Sigma}^D_i$. Hence, ${\Sigma}^D_i$ is now contractible since $\text{Int}(\Sigma^{D}_i, \gamma^1)=0$, rendering the topological operator on the right-hand side the identity.  The graphic idea is shown in Figure \ref{Fig4}.

\bbox[title=Another definition of Symmetry]
A symmetry in a quantum field theory in $D+1$-dimensions is the action of a topological operator $\hat{U}_g(\Sigma^D)$ of dimension $D$.\\

In the presence of a global continuous symmetry, there is a conserved charge over time. A charged operator $\hat{\phi} \in {\cal H}$ transforms under a symmetry operator $\hat{U}:{\cal H}\rightarrow {\cal H}$ if it carries the associated global charge,  meaning that the symmetry transformation is generated by the global charge $Q$,
\be
\hat{U}_g(\Sigma^D_i)=\e\left(i\alpha \hat{Q}_i\right),
\ee
acting on the quantum field or {\it charged operator}  $\hat{\cal O}$ as
\be
\hat{U}_g \O=R(g) \O\hat{U}_g^{-1}.
\ee
The symmetry operator $\hat{U}$ transforms a field state $\O$ surrounded by $\widetilde{\Sigma^D_i}$ to a state $\O '$ outside $\widetilde{\Sigma}^D_i$, allowing $\widetilde{\Sigma}^D_i$ to shrink into a point. \\
\ebox

For a gauged symmetry, the  electric charge $q^{(e)}_i$ is computed by integrating the current over a compact space $\widetilde{\Sigma}^D_i$, which can be extended to an integral over the whole space-time by replacing\footnote{This is actually the so-called minimal coupling in classical and quantum mechanics.}
\be
\PD(\widetilde{\Sigma}^D_i ) \longrightarrow \PD(\widetilde{\Sigma}^D_i) + A_1,
\ee
implying that
\be
q^{(e)}_i=\int_{\widetilde{\Sigma}^D_i}\ast J_1= \int_{S^{D-1}_i}\ast F_2 \longrightarrow \left(\text{Link}(S^{D-1}_i, \gamma^1)+ \int_{\gamma^1} A_1\right)\int_{S^{D-1}_i}\ast F_2,
\ee
reflecting that there is an interaction between the current $J_1$ and the gauge field $A_1$.  We have also added the Linking number $\text{Link}(\gamma^1, S^{D-1})$ counting the number of $0$-dimensional objects carrying an electric gauge charge. Therefore, the symmetry operator is now given by
\be
\hat{U}^{\text{gauge}}_g(S^{D-1}_i)=\e \left(-iq^{(e)}_i \text{Link}(\gamma^1,S^{D-1)})\right),
\ee
and acts on a charged operator $W(\gamma^1)\O$ where the Wilson line $W(\gamma^1)$ is defined as
\be
W(\gamma^1)=\text{exp}\left(iq^{(e)}_i\int_{\gamma^1}A_1\right).
\ee
Notice that $\text{Link}(S^{D-1}_i,\gamma^1)$ counts the number of Wilson lines inside $S^{D-1}_i$, but if one measures the total electric charge $q_{T,i}^{(e)}$ without knowing the number of particles on which the charge has been distributed, then this electric charge is the one that counts the number of Wilson Lines inside $S^{D-1}_i$. \\

In a similar way, from the magnetic dual point of view, one can define the {\it t'Hooft Operator} as
\be
T(\Sigma^{D-2})=\e\left(i q^{(m)}_i\int_{\Sigma^{D-2}} \widetilde{A}_{D-2}\right),
\ee
where $\ast F_2=d\widetilde{A}_{D-2}$. The total magnetic charge at a given time $t_i$ counts the number of t'Hooft's lines inside $S^{D-2}$. It is easy to see that these operators form a symmetry group where the operators are generated by the charge operators $q^{(e)}_i$ and $q^{(m)}_i$.\\ 

Finally, for a broken symmetry, i.e., for $d\ast J_1\ne 0$, we have two perspectives: 
\begin{enumerate}
\item
At some time $t_f>t_i$, $\gamma^1$ has a boundary $\pi^0$ for which ${\Sigma}^D_f$ is contractible to a point, and in consequence $\text{Int}(\Sigma^{D}_f, \gamma^1)=0$. The symmetry operator is then the identity $\hat{I}$, this is
\be
\hat{U}_g(\Sigma^D_f)=\exp\left(i \alpha ~\text{Int}(\Sigma^D_f, \gamma^1)\int_{\Sigma^D_i}\ast J_1\right)= \hat{I}.
\ee
\item
Let us consider that the symmetry is broken by the presence of a particle with global charge $q$, localized in time and space, such that the current is not closed\footnote{This is, $d\ast J_1=q \,\delta^{(D+1)}(x^{\mu}-x_0^{\mu})d(Vol)$, where $\delta^{(D+1)}(x^{\mu}-x_0^{\mu})$ is Dirac's delta distribution.}
\be
\int_{\W} d\ast J_1= q.
\ee
Then the intersection number is not zero but has different values at time $t_i$ and time $t_f$. Define the relative intersection number $\text{I}_R$ at time $t\in[t_i,t_f]$ as
\be
\text{I}_R(\Sigma^D_t, \gamma^1)=\text{I}(\Sigma^D_f, \gamma^1)-\text{I}(\Sigma^D_t, \gamma^1),
\ee
and the corresponding symmetry operator as
\be
\hat{U}_g(\Sigma^D_t)=\exp\left(i \alpha~ \text{I}_R(\Sigma^D_t, \gamma^1)\int_{\Sigma^D_t}\ast J_1\right).
\ee
Therefore,  since $\text{I}_R(\Sigma^D_f,\gamma^1)=0$, in terms of the relative global charge, we can also associate the identity to the symmetry operator at time $t=t_f$.\\
 \end{enumerate}


\begin{tcolorbox}[title=Why charge is quantized? Dirac's quantization, label=dirac, breakable]
As previously mentioned, at the quantum level, charge is quantized implying that it takes values over the integers $\Z$. This follows from various results in Quantum Mechanics and Quantum Field Theory. We present here Dirac's argument, which basically states that if a magnetic charge exists, then electric charge must take discrete values. The argument is as follows.\\

Take the  operator $\hat{U}^{\text{gauge}}_g(S^{D-1}_i)$ acting on a closed trajectory on a charged operator ${\cal O}_{\gamma^1}$ carrying some gauged charge (electric and magnetic). Hence
\be
\hat{U}^{\text{gauge}}_g(\Sigma^D_i)\hat{\cal O}_{\gamma^1}=\e\left(i\alpha q^{(e)}_i \oint_{\gamma^1} A_1\right) \Oo,
\ee
where $\alpha$ is the continuous infinitesimal parameter. Since for a non-global exact 1-form $A_1$,
\be
\oint_{\gamma^1} A_1= \oint_{S^2} F_2= q_m,
\ee
we have that
\be
\alpha q_e q_m= 2\pi n, \qquad n\in\Z.
\ee
This is Dirac quantization which leads us to consider the abelian free part of cohomology to be $\Z$. Notice that we have assumed all the necessary conditions to have a magnetic charge, as described previously. This implies that we have a topologically nontrivial $U(1)$-bundle, regarding the charge operators (or quantum fields) to be sections of a line bundle. 
\end{tcolorbox}
\vspace{0.5cm}

\section{Generalized symmetries and Cobordisms}
Generalized symmetries have captured the attention of a considerable part of the theoretical physics community over the last few years. Here, we address this generalization using the formalism presented in the previous section, where the usual symmetries in quantum field theory (QFT) can be represented as differential forms. We invite the reader to consult references \cite{Banks:2010zn, Gaiotto:2014kfa} and \cite{Luo:2023ive, Bhardwaj:2023kri, Apruzzi:2023uma, Brennan:2023mmt, Schafer-Nameki:2023jdn, Bhardwaj:2023ayw, Benedetti:2023ipt, Bhardwaj:2023wzd, Gomes:2023ahz, Cordova:2022ruw, Benedetti:2022zbb, Sharpe:2015mja}, where we have included some (but not all) interesting reviews on the topic. We also discuss some relations to cobordisms \cite{miller1994notes}, motivated by recent studies in physics initiated in \cite{McNamara:2019rup}. Some excellent reviews and applications can be found in \cite{Makridou:2023wkb, Blumenhagen:2022bvh, Blumenhagen:2022mqw, Andriot:2022mri, Blumenhagen:2021nmi, Dierigl:2020lai, Montero:2020icj}. \\

\subsection{Generalized $p$-form symmetries}
The cohomological approach to symmetries enables a natural extension of the notion of symmetry to the setting of $p$-forms: let us consider a current represented by a $(D-p)$-form
\be
\ast J_{p+1} \in C^{D-p}(\W),
\ee
which can be closed or not. For the first case, we can say that it defines a class $[\ast J_{p+1}]\in\H^{D-p}(\W;\Z)$, interpreted as a $p$-form symmetry, as a generalization of the QFT symmetries we review in the last section.  Following this nomenclature, we shall refer to the usual QFT symmetries as $0$-form symmetries. This definition makes sense if one thinks of it as a symmetry that defines a global charge carried by a $0$-dimensional object, which in turn defines a current $J_1$ along a world line $\gamma^1$ and with compact support on $\Sigma^D_i$. \\

For a $p$-form symmetry, all the above concepts are easily generalized as follows. A global charge $Q_i^\p$ related to a $p$-form symmetry is defined as
\be
Q_i^\p=\int_{\Sigma_i^{D-p}}\ast J_{p+1},
\ee
with  $Q_i^{\p} \in \text{Hom}(\H^{D-p}, \H_{D-p})$, rendering it topological. The charge, when non vanishing, is carried by a $(p+1)$-dimensional object. By Poincar\'e duality, this defines  a world volume $\gamma^{p+1}$ class in $\H_{p+1}(\W;\Z)$.\\

\begin{figure}[h]
  \centering
  \includegraphics[width=0.5\textwidth]{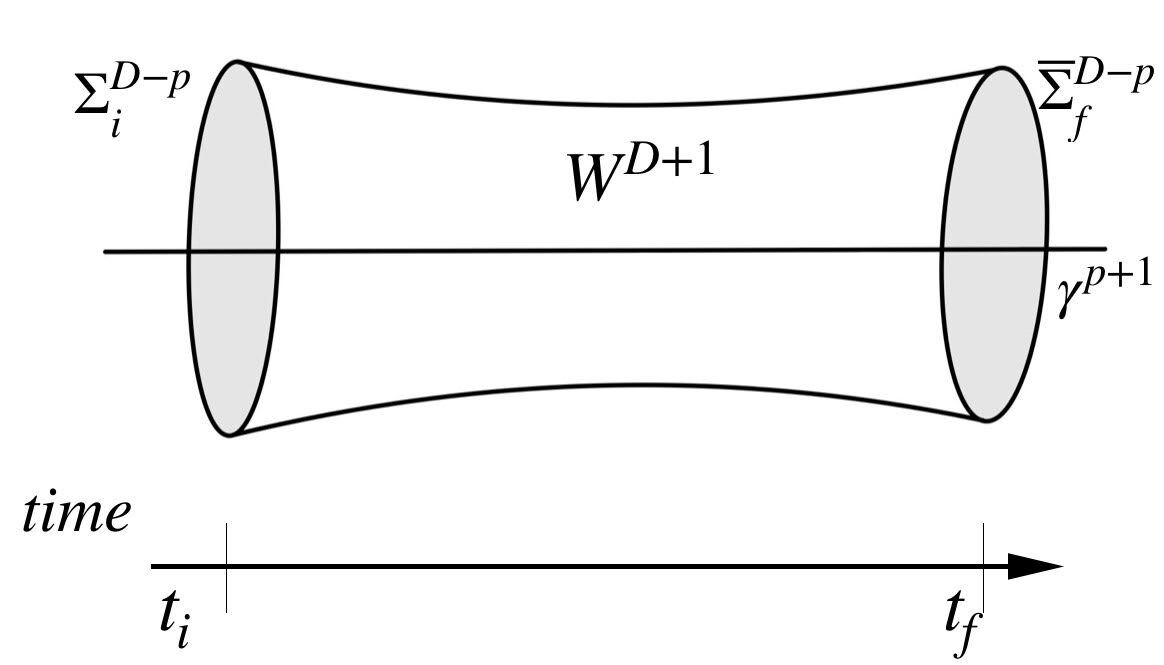}
  \caption{Conservation of generalized charge.}
  \label{Fig5}
\end{figure}

A conserved vanishing global charge $Q^\p_i$ follows from zero classes $[\ast J_{p+1}]\in\H^{D-p}$ and/or $[\Sigma_i^{D-p}]\in\H_{D-p}$. As in the previous case, having an exact $(p+1)$-form current $\ast J_{p+1}=d\ast F_{p+1}$ implies that we can associate a world-volume $\gamma^{p+1}$ with a boundary $\partial\gamma^{p+1}=\pi^p$ such that the spatial cycle $\Sigma_i^{D-p}$  contracts into a point, making it clear that there is a vanishing global conserved charge $Q_i^\p$. All the mathematical structures needed to properly define a gauge field on the base manifold $\Sigma^{D-p}_i$ are also assumed. See Figure \ref{Fig5}.\\

\begin{figure}[h]
  \centering
  \includegraphics[width=0.5\textwidth]{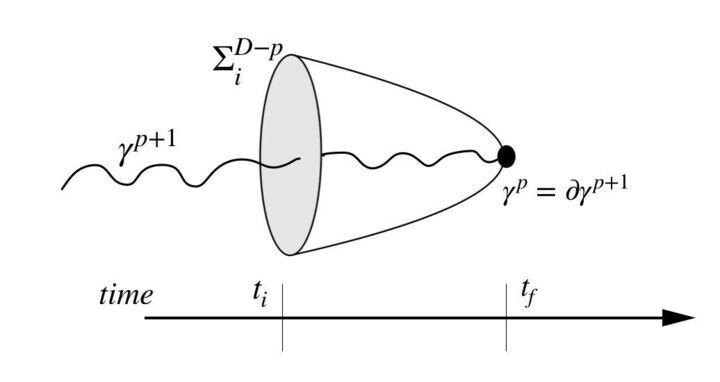}
  \caption{A broken $p$-form symmetry.}
  \label{Fig6}
\end{figure}

For the case in which a global $p$-symmetry is broken, we have that 
\be
d\ast J_{p+1}=\ast J_p,
\label{gsymm}
\ee
for a non trivial $\ast J_{p} \in C^{D-p+1}$ defining a zero class in $\H^{D-p+1}(\W;\Z)$. It is now obvious that a generic expression like $d\ast J_{p+1}=\ast J_{p}$ can be interpreted in two different—but equivalent—ways,  by identifying the global current as the differential form in the lhs or in the rhs of expression \ref{gsymm}, i.e., we can say that a $p$-symmetry has been broken by the current $J_p$ or rather that a $(p-1)$-symmetry has been gauged. See Figure \ref{Fig6}.\\

\begin{tcolorbox}[title=A gauged symmetry is a broken global symmetry, label=mr2]
 A gauged $p$-form symmetry on a $p$-dimensional space $S^p$ is equivalent to a broken global $(p+1)$-symmetry on $(p+1)$-dimensional space $\widetilde{\Sigma}^{p+1}$, with $\partial \widetilde{\Sigma}^{p+1}=S^p$.
 \end{tcolorbox}
 
 More results follow from generalizing a symmetry:
 \begin{enumerate}
 \item The $p$-dimensional objects carrying a global charge $Q_i^\p$ can be identified with  D$p$-branes in string theory. Therefore,  string theory seems compatible with generalized symmetries.
 \item We can generalize the corresponding Lagrangian in the case of an abelian  gauged $p$-symmetry as
 \be
 \L= F_{p+2}\wedge\ast F_{p+2} + A_{p+1}\wedge \ast J_{p+1},
 \ee
 \item It is possible to generalize the Wilson line and t'Hooft operators as
 \begin{eqnarray}
 W^\p(\gamma^{p+1})&=&\e\left(iq^{(p)}_e\int_{\gamma^{p+1}}A_{p+1}\right), \nonumber\\
 T^\p(\Sigma^{D-p-2})&=&\e\left(iq^{(p)}_m\int_{\Sigma^{D-p-2}}\widetilde{A}_{D-p-2}\right),
 \end{eqnarray}
 with similar interpretations.\\
 \end{enumerate}

\begin{tcolorbox}[title=Electric charge in 4-dimensions as a 1-form global symmetry breaking, label=maxwellbroken, breakable]
The inhomogeneous Maxwell's equations in $D=3$  in the presence of sources as the electric density charge $\rho$ or the electric current $\vec{J}$, can be written in terms of differential forms as  $\ast J_1=d\ast F_2$. Our first interpretation is that the 0-form symmetry described by the current $J_1$ has been gauged. However, seen from the perspective of a 1-form symmetry, where the associated current is actually a $2$-form as $\ast F_2$, we can say that the presence of sources break the global 1-form symmetry.\\

In the absence of sources, the $2$-current  $\ast F_2$ satisfies $d\ast F_2=0$. We have a 1-form global symmetry and in consequence a conserved global charge given by
\be
Q^{(1), (e)}_{i}=\int_{{\Sigma}^2_i} \ast F_2 \in \Z.
\ee
Notice that this is NOT the electric gauged charge defined in \ref{electricq}. This is actually an electric global charge generating a $U(1)$-symmetry and computed by integration on the whole space $\Sigma^2_i$ at some time $t_i$. Similarly we can define a magnetic global charge operator by
\be
Q^{(1), (m)}_i=\int_{\Sigma^2_i} F_2 \in \Z,
\ee
which make sense as a magnetic dual for $D=3$. For another dimensionality, magnetic-duals involve objects of different dimensions.  The group of 1-form symmetries is then $U(1)^e\times U(1)^m$. By adding charged matter this symmetry is broken. We are interested in understanding what is the meaning of breaking this 1-form symmetry, since adding charged matter gives us Maxwell's equations with sources.\\

From the point of view of the electric $U(1)^{(e)}$-form, the current $\ast F_2$ is not longer closed, this is $d\ast F_2=\ast J_1$. By Poincar\'e duality we can say that the corresponding cycle $\gamma^2$ has a boundary $\gamma^1\in C_1(\W)$ located at time $t_f>t_i$. Therefore, the cycle $\Sigma^2_i$ shrinks into a point. As mentioned, this means in the QFT perspective that the symmetry operator is actually the identity, this is
\be
\hat{U}_g^{(e)}(\Sigma^2_f)=\e\left( i\alpha Q^{(1), (e)}_f\right)=\hat{I}.
\ee
Since breaking the global symmetry means that at some time $t_i$ there is a non-zero integral-valued global charge, $Q^{(1),(e)}_i=n\in\Z$, we have that $n\alpha=2\pi \kappa$ for some $\kappa\in \Z$. Therefore
\be
\alpha=\frac{2\pi \kappa}{n},
\ee
implying that the continuous group $U(1)$  is broken to $\Z_n$. The presence of charged matter in Maxwell's equation is then interpreted as the breakdown of a 1-form symmetry. Hence, the 0-form gauged symmetry in Maxwell's equations is actually a 1-form global broken symmetry.\\
\end{tcolorbox}

\subsubsection{Generalized gauged symmetries}
Gauged symmetries have an important interpretation in the context of {\it String Theory}. In this case, the current $\ast J_{p+1}$ is carried by a $(p+1)$-dimensional object. For an exact $(D-p)$-form $\ast J_{p+1}=d\ast F_{p+2}$, the global charge $Q^\p_i$ vanishes, but we can still define an electric charge given by
\be
\mu^{\p,e}_i=\int_{S_i^{D-p}}\ast F_{p+2},
\ee
and a magnetic charge
\be
\mu^{\p,m}_i=\int_{S_i^{p+2}} F_{p+2}.
\ee
These $p$-dimensional objects are known in String Theory as D$p$-branes and are fundamental objects along with quantum strings. Essentially, they are higher dimensional non-perturbative dynamical objects on which the ends of open strings end, fulfilling Dirichlet boundary conditions\footnote{This is the reason for the name D(irichlet)-branes.}. The quantum oscillations of such strings are the source of the gauged fields carrying the gauge forces, such as electromagnetism, and the nuclear strong and weak forces.\\

The exactness of the current $\ast J_{p+1}$ also follows from the variation in the Lagrangian given by
\be
\L= F_{p+2}\wedge\ast F_{p+2} + C_{p+1}\wedge\ast J_{p+1},
\ee
where $F_{p+2}=dC_{p+1}$ and $C_{p+1}$ are known as the Ramond-Ramond (RR) potentials. Their construction from the quantum dynamics of the open string constrains the values for $p$, implying the existence of two types of string theories called type IIA and type IIB. The former has even values for $p$, while the latter has only odd values.\\

\subsection{Cobordisms}
We have seen that, in the presence of 0-form global continuous symmetries, the space cycles $\Sigma^D_{i,f}$ over which we measured the global charge belong to the same homology class. This is trivially generalized to cycles $\Sigma^{D-p}_{i,f}$ for $p$-form symmetries. However, it seems plausible that two cycles must be related to each other by an equivalence relation, even if they possess more general structures, such as being a base manifold with orientability and spin, or $\spin$ structures. Therefore, it is desirable to relate such spaces by more general equivalence relations that consider the presence of extra mathematical structures. This is the reason we are introducing {\it cobordisms}.\\

%
%

Let $\Sigma_i^q$ be a $q$-dimensional space-like submanifold of $\W$ with a Euclidean metric and a generic structure $G$ on it. We say that $\Sigma_i^q$ is cobordant to $\Sigma_f^q$ if there exists a $(q+1)$-dimensional manifold $\Pi^{q+1}$ such that $\partial\Pi^{q+1}=\Sigma_i^q\sqcup\overline{\Sigma}_f^q$, where $\overline{\Sigma^q_f}$ is oriented oppositely compared to $\Sigma^q_f$. It is also possible that $\Sigma_i^q$ and $\Sigma_f^q$ can, in turn, be composed of a disconnected sum of more spatial regions. As the reader can see, this is exactly the definition we have used for $\Sigma^{D-p}_i$ as the spatial region to measure a global charge $Q^\p_i$, and $\Pi^{q+1}$ is the transverse space to $\gamma^{p+1}$ in $\W$.\\

It is easy to check that being cobordant is an equivalence relation since 
\begin{itemize}
\item 
every $\Sigma_i^q$ is cobordant to itself.
\item
If $\Sigma_i^q$ is cobordant to $\Sigma_f^q$, then (in our case, by reversing time), $\Sigma_f^q$ is cobordant to $\Sigma_i^q$.
\item
If $\Sigma_i^q$ is cobordant to $\Sigma_f^q$ and $\Sigma_f^q$ is cobordant to $\Sigma_F^q$, then $\Sigma_i^q$ is cobordant to $\Sigma_F^q$.
\end{itemize}

This allows for the definition of a cobordant equivalence class by
\be
[\Sigma_i^q]=\large\{ \Sigma_f^q| \Sigma_i^q\sim\Sigma_f^q\large\},
\ee
where  $\Sigma_f^q$ is the space slice used to measure the charge $Q_f^q$ at some time $t_f>t_i$. Notice that causality is not assumed. This set of equivalence classes establishes a $q$-graded group, called the $q$th-Cobordism{\it group} $\Omega^G_q$, where $G$ denotes all the required structures on the cobordant spaces $\Sigma^q$. The graded cobordism group is then given by
\be
\Omega^G=\bigoplus_{q=0}^{D} \Omega^G_q,
\ee
where the group operation is defined through the disjoint union of manifolds $\Sigma^q$. Then, we have
\begin{itemize}
\item
The zero element $[0]$ is defined as $[\Sigma_i^q\sqcup {\Sigma}_i^q]$. For our case, notice that the corresponding global charge measured through this opposite-oriented manifold is negative with respect to the original one.
\item
The sum operation is defined as $[\Sigma_i^q]+[\Pi_i^q]=[\Sigma_i^q\sqcup\Pi_i^q]$. Notice that in physics, we require the two space manifolds to be defined at the same time (this resembles the expectation values considered in QFT).
\end{itemize}

It is important to distinguish between two different physical cases. If the space foliation we are taking to compute the global charge at time $t_i$ has no boundary, i.e. $\partial\Sigma^q_i=0$ we shall denote the cobordism group as $\Omega^G_q$. If we consider a compact space foliation $\widetilde{\Sigma}^q_i$ such that  $\partial\widetilde{\Sigma}^q_i=S^{q-1}_i$, we shall denote the cobordism group as $\widetilde{\Omega}^G_{q-1}$. Notice that zero classes in $\Omega^G_q$ are not necessarily zero classes in $\widetilde{\Omega}^G_{q-1}$. Hence, in terms of cobordism classes, 
 the following assertions hold:
\begin{enumerate}
\item
For a global continuous symmetry, $\Sigma^{D-p}$ defines an equivalence class in $\Omega^G_{D-p}$ with the cobordism given by the transversal space to $\gamma^{p+1}$ in $\W$, which we call ${\cal W}^{D-p}$. 
\item
If the global $(p)$-form symmetry is broken, we have seen that $\Sigma^{D-p}$ shrinks into a point. At the level of cobordism classes, this means that $[\Sigma^{D-p}]=[0]\in \Omega^G_{D-p}$, which also describes a gauged symmetry with the gauged electric charge measured on $S^{p-1}$.\\
\item
The global continuous conserved charge $Q_i^{(p)}$  is still an isomorphism between cohomology and homology, but as in the case of a principal vector bundle, we require constructing a map from the group of cobordism to cohomology. For that, it is necessary to construct an isomorphism between the oriented cobordism group of $(D-p)$-dimensional manifolds $\Sigma^{D-p}$ and the cohomology groups of certain spaces called {\it Thom Spaces} $T(E)$ associated with the oriented vector bundle $E$ over $\Sigma^{D-p}$, where the Thom space is the total space of $E$ modulo its zero section. The existence of this isomorphism is asserted by the Thom Isomorphism theorem. This is
\be
\Omega^G_{D-p}\cong \H^\bullet (T(E)),
\ee
where $G$ denotes the structures we are assuming as the orientation ($SO$),  \text{spin} , and $\spin$. This theorem provides a bridge between cobordism and cohomological invariants associated with vector bundles such as the global charge. Therefore, the global conserved charge is an integer value computed from the isomorphisms between  $\Omega^G_{D-p}$ and $\H^{D-p}(\W;\Z)$.
\item
Gauged symmetries and broken higher symmetries are, in this context, represented by zero classes in the cobordism group $\Omega^G_q$, although by non-zero classes in $\widetilde{\Omega}^G_{q-1}$. This is actually quite important in physics. It seems that gauge symmetries, which play an important role in particle physics, are encoded in zero classes in cobordism groups. Non-zero classes are, on the other hand, associated with non-zero conserved global charges.
\end{enumerate}

%
%
%
%

\subsection{General sources for global and gauged charges}

Since we are allowing the presence of higher $p$-field forms, it is possible to define charges in another way, through the interaction with {\it fluxes} that we proceed to define. Consider the compact spacetime $\W$ of the form
\be
\W= \Sigma^{D-p}\times\gamma^{p+1}.
\ee
By {\it K\"unneth formula} , we have that
\be
\H^n(\W)=\bigoplus_{r+s=n}\left(\H^r(\Sigma^{D-p})\otimes \H^s(\gamma^{p+1})\right),
\ee
where we have identified $\gamma^{p+1}$ as a manifold. We define {\it a flux} as a $r$-form $\omega_r$ defining a class $[\omega_r] \in \H^r(\Sigma^{D-p})$. Being closed, $\omega_r$ can be identified with a flux without associated sources.\\


Therefore, since there are different fluxes for a given  $q$, it is possible to define the charge throughout a product of different forms and fluxes.  The explicit way of constructing a $(D-p)$-form  (emulating a current $\ast J_{p+1}$ ) from which a global charge can be computed establishes different terms in a corresponding Lagrangian. Hence, a way to compute such charge (up to two terms) is
\be
Q^\p_i=\int_{\Sigma^{D-p}}H_{q}\wedge \tilde\ast F_q=\int_{\Sigma^{D-p}}H_q\wedge \widetilde{F}_{D-p-q},
\ee
where $\tilde\ast$ is the Hodge dual operator restricted to the subspace $\Sigma^{D-p}$ 
\be
\tilde\ast: \H^q(\Sigma^{D-p})\longrightarrow \H^{D-p-q}(\Sigma^{D-p}),
\ee
and $H_q$ is a flux in $\H^q(\Sigma^{D-p})$. This, in turn, defines new terms in a Lagrangian. The goal of this section is to understand the construction of such terms from the point of view of cohomology.\\

\subsubsection{Charges by fluxes}
Consider the case in which we have a $q$-form $H_q \in \H^q(\Sigma^{D-p}_i;\Z)$ and a $(D-p-q)$-form $F_{D-p-q}\in \H^{D-p-q}(\Sigma^{D-p}_i;\Z)$.  By Kunneth's formula, the charge is defined as
\be
Q^\p_{i,1}=\int_{\Sigma^{D-p}}H_q\wedge F_{D-p-q}.
\ee
Also, if there is a $p$-dimensional object carrying the charge, we have that 
\be
Q^\p_{i,2}=\int_{\Sigma^{D-p}} \ast J_{p+1},
\ee
from which it is said that $[\ast J_{p+1}]=[H_q\wedge F_{D-p-q}]$. This means we can also measure the global charge not from an object carrying it, but from a configuration of  fluxes.\\

However, something very important is happening here. Notice that there are two sources contributing to the global charge $Q^\p_i$: the $p$-dimensional object (D$p$-brane in the context of string theory) and the fluxes $H$ and $F$. Therefore, the total charge at time $t_i$ is given by
\be
Q^\p_{i, T}=\int_{\Sigma^{D-p}}\large(\ast J_{p+1}+H_q\wedge F_{D-p-q}\large).
\ee
Now, if the sum of the current and fluxes defines an exact form in $\H^{D-p}$, the global charge vanishes, from which 
\be
d\ast F_{p+2}=\ast J_{p+1}+ H_q\wedge F_{D-p-q},
\ee
indicating the existence of a Lagrangian of the form
\be
\L=F_{p+2}\wedge\ast F_{p+2} - C_{p+1}\wedge\left(\ast J_{p+1} + H_q\wedge F_{D-p-q}\right).
\ee
These kinds of terms are part of field Lagrangians in {\it Supergravity theory}. The construction of a supergravity theory seems to be consistent with assigning the spaces on which we evaluate global symmetries to the trivial (zero) cobordism class.\\

\begin{tcolorbox}[title=Supergravity in 10-Dimensions, label=SUGRA, breakable]
Supergravity (SUGRA) is a theory of local (gauged) supersymmetry  formulated in various dimensions. In particular for $D=9$, it represents the low effective limit of the field content arising from the massless sector of string theory.  SUGRA equations of motion are of the form
\be
dF_{p+2}= H_3\wedge F_{p}
\ee
with $F_{p+2}=d C_{p+1}$. In the presence of D$p$-branes, the equations are written as
\be
dF_{p+2}= H_3\wedge F_{p} + \ast J_{7-p}.
\ee
The fluxes fulfill the so called Bianchi identities, 
\be
dH_3=0, \qquad dF_{p}=0,
\ee
or equivalently, they define classes in $\H^3(\W;\Z)$ and $\H^p(\W;\Z)$ respectively.\\

In type IIB string theory, $D=9$, and a D3-brane charge is actually an electric charge from a gauged 3-form symmetry. We can write
\be
\ast J_4= d\ast F_5, \qquad \text{and} \qquad H_3\wedge F_3=d\ast F_5,
\ee
where $H_3$ and $F_3$ are field forms arising from the closed string massless state. A D3-brane charge comes from a flux configuration given by $H_3$ and $F_3$ with $H_3$ supported on a transversal cycle to $\gamma^4$ in $\W$.
Equivalently we can say that a 4-form symmetry has been broken by the presence of D3-branes or by the presence of fluxes.  
Can we describe a transition between these two sources?\\
\end{tcolorbox}

\subsubsection{Topological transition between fluxes and branes}

Let us assume a simple case in which we have a $p$-form global symmetry and a conserved charge $Q^\p_i$. The current $\ast J_{p+1}$ is closed, and the actual value of the charge is determined by the cohomology class $[\ast J_{p+1}]$. But, as we have discussed, it is also possible to get the exact same value by turning on the appropriate fluxes, such that
\be
[\ast J_{p+1}]=[H_q]\cdot [F_{D-p-q}].
\ee
Hence, we have at some time $t_i$ a global conserved charge ($d\ast J_{p+1}=0$) carried by a $p$-dimensional object with a world-volume $[\gamma^{p+1}]\in\H_{p+1}(\W;\Z)$ such that $\partial\gamma^{p+1}=0$. However, at some time $t_2$, the same charge is measured but now from a set of fluxes $H_q$ and $F_{D-p-q}$ supported on appropriate cycles.  Although the charge is conserved, there must be an explanation of how we can change the sources of the charge. There must be a topological transition between the $p$-dimensional object and the set of fluxes. This transition was studied  in \cite{Maldacena:2001xj} where it was shown that it  is driven by an instantonic brane described in terms of a refinement of cohomology through the Atiyah-Hirzebruch Spectral Sequence.\\


\section{Second generalization: $p$-form 3-symmetries by K-theory}
In this section, we make use of some standard results in string theory and establish a relation to K-theory. Once again, the number of references is substantial. Some well-known books and articles include \cite{Minasian:1997mm, Witten:1998cd, Bouwknegt:2000qt, Diaconescu:2000wz, Diaconescu:2000wy, Witten:2000cn, Maldacena:2001xj, Moore:2003vf, Evslin:2006cj, Grady:2019man}.\\

Within the context of string theory, the field strengths arising from the quantization of the massless states can be nicely divided into different sectors. In the case of a closed string, these two sectors refer to the chirality of the spinors on the world-sheet and are named Ramond-Ramond (RR) and Neveu-Schwarz-Neveu-Schwarz (NS-NS). All fields in the RR sector have an associated gauged charge carried by D-branes. In the NS-NS sector, we have another flux represented by a 3-form flux, called the NS-NS 3-form flux $H_3$. In this section, we shall use this flux to induce D-brane charge through fluxes. For instance, a D$p$-brane has an associated gauge charge $\mu_p^e$ given by
\be
\mu_p^e=\int_{S^{D-p-1}_i} \ast F_{p+2},
\ee
but also the flux configuration $H_3\wedge F_{D-p-3}$ contributes to the charge, driving a transition from D-branes into fluxes. This follows from generalizing cohomology through the use of the flux $H_3$. This simple ansatz will connect to a generalized cohomology version called K-theory. Notice that all we shall describe here is valid up to the presence of the flux $H_3$. This, as studied in the last two decades, is necessary to have a correct description of string theory at the effective level (moduli stabilization).\\

\subsection{The Atiyah-Hirzebruch Spectral Sequence}

Assuming the presence of $H_3\in \H^3(\W;\Z)$, it is possible to achieve a refinement of cohomology through the Atiyah-Hirzebruch Spectral Sequence (AHSS) which is a tool for computing a generalized cohomology theory, such as $K$-theory, starting from the ordinary cohomology of a space.\\

The spectral sequence begins with
\be
E_2^{p,q}\cong H^p\bigl(X,K^q(\mathrm{pt})\bigr),
\ee
and evolves through successive pages
\be
E_2 \Rightarrow E_3 \Rightarrow E_4 \Rightarrow \cdots \Rightarrow E_\infty,
\ee
with differentials
\be
d_r:E_r^{p,q}\longrightarrow E_r^{p+r,q-r+1}.
\ee

For complex $K$-theory,
\be
K^q(\mathrm{pt})=
\begin{cases}
\mathbb Z, & q \ \text{even},\\
0, & q \ \text{odd},
\end{cases}
\ee
so the initial data are essentially the integral cohomology groups $H^p(X,\mathbb Z)$. At each stage, the next page is obtained as
\be
E_{r+1}^{p,q}
=
\frac{\ker d_r}{\operatorname{im} d_r}.
\ee
Therefore, the differentials eliminate cohomology classes that cannot be lifted to genuine $K$-theory classes.
To understand the limiting page $E_\infty$, consider a CW decomposition
\be
X^0\subset X^1\subset \cdots \subset X^n\subset \cdots \subset X,
\ee
where $X^n$ denotes the $n$-skeleton of $X$. This induces a filtration of the $K$-theory group,
\be
K(X)=F^0K(X)\supseteq F^1K(X)\supseteq F^2K(X)\supseteq\cdots,
\ee
where
\be
F^nK(X)
=
\ker\left(
K(X)\longrightarrow K(X^{n-1})
\right).
\ee

Thus, $F^nK(X)$ consists of those $K$-theory classes that become trivial when restricted to the $(n-1)$-skeleton of $X$. Equivalently, these are classes that require cells of dimension at least $n$ to be detected.\\

The limiting page of the AHSS is related to this filtration through
\be
E_\infty^{p,q}
\cong
\frac{F^pK^{p+q}(X)}
     {F^{p+1}K^{p+q}(X)}.
\ee

Hence, $E_\infty$ does not, in general, give $K(X)$ directly. Instead, it gives the associated graded group
\be
\operatorname{Gr}K(X)
=
\bigoplus_p
\frac{F^pK(X)}{F^{p+1}K(X)}.
\ee
Recovering the full $K$-theory group from $E_\infty$ may require solving extension problems.\\

In string theory, this construction is particularly useful for classifying D-brane charges. Ordinary cohomology provides a first approximation to the possible charges, while the AHSS differentials determine which cohomology classes survive as genuine $K$-theory classes. In particular, classes that are not in the kernel of a differential, such as $d_3$, are removed, while classes differing by elements in the image of a differential are identified.\\

The graded group $\operatorname{Gr}K(X)$ is isomorphic to $K(X)$ (the K-theory group) if the exact short sequence
 \be
 \begin{tikzcd}
 0\arrow[r, ""] & K_p\arrow[r, ""]&E^p_r\arrow[r,""]&K_{p+1}\arrow[r,""]&0,
 \end{tikzcd}
 \ee
 is trivial for all $p$ with $K_p=\ker{d^p_r}$. In our case, since we are not considering torsion components in cohomology, this is always true\footnote{ If torsion is not considered, we have almost a perfect match between cohomology and K-theory, and in the presence of $H_3$, the twisted cohomology given by the first step in the AHSS equals the twisted K-theory group.  However,  string theory allows the presence of extra high-dimensional objects called {\it orientifolds} in order to reduce the number of supersymmetric generators in the effective theory, which is a sensible proposal since, at the effective level, we have a structure given by the standard model of particles, compatible at most with a single supersymmetric generator. Mathematically, this translates into the presence of torsion components in the cohomology groups, altering as well the construction of the AHSS. We are not considering such cases in these lectures. Some implications of the presence of torsion have been studied by one of the authors: \cite{Loaiza-Brito:2003ivs, Loaiza-Brito:2004ajy,  Loaiza-Brito:2007umh, Garcia-Compean:2013sla, Damian:2019bkb}.}.\\

 Since, in the context of string theory, there is not a 5-form field from the NS-NS sector,  let us describe the second order approximation given by the differential map $d_3$. This is, in general, given by
 \be
 d_3: \H^p(\W;\Z)\longrightarrow \H^{p+3}(\W;\Z), 
\ee
with
\be
d_3(\omega_p)= Sq^3(\omega_p) + H_3\wedge \omega_p,
\ee
where $Sq^3$ is the {\it Steenrod square}  related to the third Stiefel-Whitney class (or equivalently, the Bockstein map of the second Stiefel-Whitney class) such that $Sq^3(\omega_p)$ vanishes for a $p$-form defined over a $\spin$ manifold. As previously discussed, we concentrate on such cases since we are interested in describing the interaction between gauge fields and fermions.\\

Under such conditions, the differential map $d_3$ reduces to
$d_3\omega_p=H_3\wedge\omega_p$. Notice that $d_3^2=0$ leads us to the definition of the group 
 \be
 E_2^p(\W;\Z)=\{\text{ker } d_3\}/\{\text{Im} d_3\},
 \ee
 consisting of all $p$ forms in $H^p(\W;\Z)$ such that $d_3\omega_p=0$, but not of the form $\omega_p=d_3\sigma_{p-3}$ for $\sigma_{p-3}\in\H^{p-3}(\W;\Z)$. This is actually a refinement of cohomology since it takes closed forms that belong to the kernel of $d_3$, which, under our conditions, means that $H_3\wedge\omega_p=0$. This last constraint is actually the {\it Freed-Witten} condition to prevent anomalies\footnote{An anomaly is a symmetry at the level of classical physics that is not preserved in the quantum version of the theory,} on D$p$-branes wrapping $\spin$-cycles.\\

 \begin{tcolorbox}[title=Freed-Witten anomaly, label=FW, breakable]
 A very simple case in which we have a Freed-Witten anomaly is as follows. Consider a D$p$-brane wrapping a cycle $\Sigma^p$ at a given time $t$ and let us consider that a flux $H_3$ has a support on $\Sigma^p$ such that
 \be
 \int_{\Sigma^3} H_3 = n\in \Z,
 \ee
 where $\Sigma^3$ is a subcycle of $\Sigma^p$. As previously discussed, we can extend the integral to the whole world-volume generated by the D-brane current by identifying $\PD (\Sigma^3)$ on $\gamma^{p+1}$ to a $p$-dimensional gauge $(p-2)$-form $A_{p-2}$, defining an effective action on the world volume $\gamma^{p+1}$ given by
 \be
 S_{\text{eff}}= \kappa \int_{\gamma^{p+1}} H_3\wedge A_{p-2}.
 \ee

 This happens since we are dealing with a gauge symmetry. By taking the variation of $S_{\text{eff}}$ with respect to $A_{p-2}$ we get the equations of motion, implying that $\kappa=0$, this is, we cannot have a non-zero flux over a D-brane. This is the Freed-Witten anomaly in its simplest form, since we have assumed that $\gamma^{p+1}$ is a $\spin$-manifold.  The anomaly can be cured if we add sources for the gauge field $A_{p-3}$, represented by a D$(p-2)$-brane ending on $\gamma^{p+1}$. \\
 
 We can also observe that it is possible to understand the Freed-Witten anomaly as a zero global charge restricted to the $(p+1)$-dimensional  space-time $\gamma^{p+1}$. Here the global charge at a given time $t_i$ is
 \be
 Q_i^{(2-p)}=\int_{\gamma^3} H_3,
 \ee
 with respect to a global $(2-p)$-form symmetry. We can say then that the global charge $Q_i^{(2-p)}$ must vanish, defining a zero cobordism class in $\Omega^G_{2-p}$ restricted to the space time $\gamma^{p+1}$ (i.e., $D=p$), with $G$ denoting the mathematical structures corresponding to an oriented and $\spin$ manifolds.
 \end{tcolorbox}

\subsection{Definition of a $p$-form $3$-symmetry.} In the following, we shall limit our description to gauged symmetries, i.e., we want to study a possible topological transition among D$p$-branes carrying gauged electric and magnetic charges and fluxes. Therefore, all currents are going to be exact: $\ast J_{p+1}=d\ast F_{p+2}$ defining a zero class in $\H^{D-p}(\W;\Z)$ and a non-zero class in $\H^{D-p-1}(\W;\Z)$. According to previous sections, this means that we have a conserved null charge and a gauged symmetry; this is $[\ast J_{p+1}]=[0] \in \H^{D-p}(\W;\Z)$ and  a non-closed form $\ast F_{p+2} \in C^{D-p-1}(\W)$.\\

 We then {\it define a $p$-form 3-symmetry} as a current $J_{p+1}$ depicted by a D$p$-brane, satisfying both
\be
d\ast J_{p+1}=0, \qquad \text{and} \qquad d_3\ast J_{p+1}=0.
\ee
This defines an equivalence class $[\ast J_{p+1}]\in E_2^{D-p}(\W;\Z)$. Although this class can be non zero in $E_2^{D-p}(\W;\Z)$  is actually the zero class in $\H^{D-p}(\W;\Z)$ since we are considering only gauged symmetries, i.e., $\ast J_{p+1}=d\ast F_{p+2}$. Then, for a $p$-form 3-symmetry, we have that $H_3\wedge \ast J_{p+1}=0$, meaning that there is no Freed-Witten anomaly associated with this D$p$-brane. \\

More follows from the definitions stated through the differential map $d$. Let us recall the physical interpretation of $[\ast J_{p+1}]$ belonging to $\H^{D-p}(\W;\Z)$. This is just, as we learned, that a D$p$-brane is stable and that the null global charge, or equivalently, the electric gauged charge, is conserved. By Poincar\'e duality, we can also interpret that  $d\ast J_{p+1}=0$ is equivalent to having a boundaryless $(p+1)$-worldvolume $\gamma^{p+1}$ defining a class in $\H_{p+1}(\W;\Z)$. We can also associate an interpretation of a current class in $E_2^{D-p}$.  Since under the map $d_3$ the class of $\ast J_{p+1}$ is sent to the zero class in $H^{D-p+3}(\W;\Z)$, we have that $H_3\wedge \ast J_{p+1}=0$ and the zero global charge cannot be measured by fluxes. A D$p$-brane does not transform into fluxes. In turn, a stable (to decay into fluxes) D$p$-brane also defines a zero class in the corresponding cobordism group.\\

\bbox[title=Stable D-branes under 3-symmetries]
A stable D$p$-brane that could transform into fluxes but does not decay into the vacuum: \\
$d\ast J_{p+1}=0 \longrightarrow [\ast J_{p+1}]=[0]\in \H^{D-p}(\W;\Z)$.\\

A stable D$p$-brane that does not transform into fluxes and does not decay into the vacuum : \\
$d_3\ast J_{p+1}=0 \longrightarrow [\ast J_{p+1}]\in E^{D-p}_2(\W;\Z)$.\\

Notice that the class in $E_2^{D-p}$ is not the zero class. In both cases, since we keep the zero class in cohomology, the spaces by which we measure the charge shrink into a point and define a zero cobordism class. \\
\be
\text{Stable D}p\text{-brane} \rightarrow [0]\in {\widetilde{\Omega}}^{G}_{D-p}.
\ee
 \ebox

It is interesting to ask whether a D-brane decays into fluxes while keeping the conservation of global and gauge charge, and if this situation still leads us to associate the current brane space with a zero class in the cobordism group. \\

\subsection{A broken or gauged 3-symmetry.} Consider then a current $\ast J_{p+1}$ related to a $p$-form 0-symmetry. It satisfies 
\be
\ast J_{p+1}=d_3\ast J_{p+4}. 
\ee
We define this as being {\it exact under $d_3$.} From the point of view of the current $\ast J_{p+4}$, we have a broken 3-symmetry since $\ast J_{p+4}$ is not closed under $d_3$. What is the physical interpretation of these two situations?\\

Let us consider a $d_3$-exact current associated with a D$p$-brane. Since we are focusing on gauged symmetries, we can also say that
\be
d\ast F_{p+2}=\ast J_{p+1}=d_3\ast F_{p+4}=H_3\wedge \ast F_{p+4},
\ee
where $\ast F_{p+4}$ is, in general, another flux (notice that it cannot be a current associated with a D-brane) and where the gauge symmetry has induced a map 
\be
d_3:\H^{D-p-3}(\W)\longrightarrow \text{H}^{D-p}(\W),
\ee
acting on the closed field strength $\ast F_{p+4}$. The charge carried by the D$p$-brane can also be computed from the fluxes $H_3$ and $\ast F_{p+4}$. We say that the D-brane has transformed into these fluxes. Hence, a D$p$-brane with an exact current with respect to the map $d_3$ is unstable to topologically transform  into fluxes, defining a zero class in $E_2^{D-p}$.\\

\begin{figure}[h]
  \centering
  \includegraphics[width=0.5\textwidth]{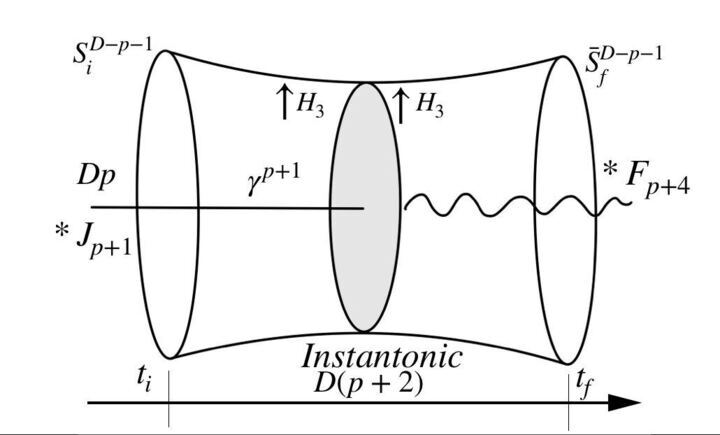}
  \caption{Topological transformation between D-branes and fluxes.}
  \label{Fig7}
\end{figure}

We have mentioned that the form $\ast F_{p+4}$ cannot be  a current associated with a D$(p+3)$-brane. This is because  we are assuming that there are D$p$-branes (for a given $p$), and by Bott periodicity from K-theory, we can conclude that there are no D$(p+3)$-branes. Hence, how can we justify the existence of a flux $\ast F_{p+4}$? This is the magnetic field strength of a D$(p+2)$-brane whose world-volume $\sigma^{p+3}$ is completely inserted in $\Sigma^{D-p}$. Hence, it is localized in time. These objects are called {\it instantonic branes} \cite{Maldacena:2001xj, Moore:2003vf, Evslin:2006cj}. See Figure \ref{Fig7}.\\

In summary, breaking the 3-symmetry means that $[d_3\ast F_{p+4}]\in \H^{D-p}(\W;\Z)$ but not in $E_2^p(\W;\Z)$. Since we are considering only gauged symmetries, this also means that

\begin{gather*}
\function{d_3}{\H^{D-p}}{\H^{D-p+3}}{0}{0} .
\end{gather*}

Therefore,  once again, the zero class plays a relevant role.  For the electric and magnetic gauged charges computed from the current of D-branes and the fluxes, there is conservation. Similarly to the case of 0-symmetries, we can relate a non-zero class in $\Omega^G_{D}$. However, unlike the scenario without fluxes, here we encounter two sources of charge. We can manipulate them to achieve a neutral electric charge by ensuring that the contributions from the fluxes offset those from the D-branes.\\

In this case
\be
\mu_p^{(e)}=\int_{S^{D-p-1}} \ast F_{p+2} -\int_{\widetilde{\Sigma}^{D-p}} H_3\wedge \ast F_{p+4} =0,
\ee
implying an equation of motion of the form
\be
d\ast F_{p+2}= H_3\wedge \ast F_{p+4},
\ee
which is naturally obtained from gauged supergravity theories or effective string theory actions in type IIB. Observe that this implies that the total electric charge on a compact $\widetilde{\Sigma}^{D-p}$ must vanish. This is known in string theory as the tadpole condition.  Hence, fulfilling the tadpole condition is compatible with a D-brane current in a zero class in $\widetilde{\Omega}^G_{D-p}$, and is followed by a zero class in ${\Omega}^G_{D-p-1}$. \\

\bbox[title=Topological transition between D-branes and fluxes, label=topo]
A stable D$p$-brane can transform into fluxes: \\
$d\ast J_{p+1}=0 \longrightarrow [\ast J_{p+1}]=[0]\in \H^{D-p}(\W;\Z)$.\\

A stable D$p$-brane that transforms into fluxes: \\
$\ast J_{p+1}=d_3\ast F_{p+4}  \longrightarrow [\ast J_{p+1}]=[0]\in E^{D-p}_2(\W;\Z)$.\\

A $Dp$-brane unstable to topologically transform into fluxes defines a zero-class in $E_2^p(\W)$ and in $\Omega^G_{D-p}$.\\
\ebox

\subsection{Implications to Quantum Gravity: the cobordism conjecture}\label{conjecture}

 Throughout these notes, we have emphasized that the conservation of gauged charges is also described by a  null cobordism class. The purpose of this emphasis is to encourage the reader to contemplate the existence of global charges induced by internal symmetries. However, the presence of globally conserved charges, associated with non-zero cobordism classes, does not appear to play a significant role in the theory, particularly once we incorporate gauge fields and their extensions in string theory. Consequently, one might question whether the presence of global charges is compatible with the additional force we have yet to consider: gravity.\\

In the last decade, there have been insights leading us to conclude that a theory of quantum gravity cannot accommodate global conserved charges (see recent reviews in \cite{Palti:2019pca, vanBeest:2021lhn, Grana:2021zvf, Agmon:2022thq}). The argument is as follows: Consider a black hole containing some amount of global charge $Q$ and some gauged (electrical or magnetic) charge $q$. While these charges exist beyond the event horizon, outside observers can indeed determine the amount of gauged charge inside the black hole, but they cannot reach the same conclusion regarding the global charge. This is because, as previously discussed, determining the gauged charge $q$ requires a closed space $S^2$ in a four-dimensional spacetime that encompasses the black hole, whereas determining a global charge $Q$ necessitates access to all space where the charge is situated. Since we cannot penetrate the horizon, the global charge remains indeterminate, and there exists no restriction on the quantity of charge within the black hole. Furthermore, this charge cannot be emitted outside the horizon via Hawking radiation since, during this process, we could potentially measure this charge at the horizon. However, due to the no-hair theorem (which stipulates that the only measurable parameters of a black hole are the gauged charge, mass, and angular momentum), this scenario is implausible. Consequently, we are left with a black hole containing an infinite number of possible quantum states that could be charged under this global symmetry. If this holds true, the entropy of the black hole would also be infinite. However, we know from the work of Bekenstein and Hawking that the entropy of a black hole is bounded by its area. Therefore, we conclude that continuous global symmetries in quantum gravity appear to be untenable. Since global symmetries are linked to non-zero classes in cobordism, it has been conjectured that a theory consistent with quantum gravity must be described by null classes in a cobordism theory, where the mathematical structures on the spacetime are yet to be fully determined \cite{McNamara:2019rup}.\\

\acknowledgments
O. L.-B. expresses his gratitude to Carolina Neira Jimenez, Diego Gallego, Sylvie Paycha, and especially to Clara Aldana for their extraordinary work as organizers of the XI Geometric, Algebraic, and Topological Methods in Quantum Field Theory in Villa de Leyva, Colombia, and for their financial support. He also thanks the other lecturers, Marcela Cárdenas, Marco Gualtieri, Kasia Rejzner, and Thomas Schick,  and the rest of the participants for creating a wonderful environment for discussions on physics and mathematics. O. L.-B. was partially supported by UG-DAIP-CIIC-2023. V. M. L.-R. was supported by a doctoral fellowship from the Secretaría de Ciencia, Humanidades, Tecnología e Innovación (SECIHTI), Mexico.
\appendix

\section{Some basics on de Rham (co)homology}
\label{apendice}
The existence of global symmetries and the corresponding conserved charges can be understood in  terms of differential forms. We shall see that many important physics concepts can be translated into simple and precise mathematical properties of  differential forms.\\

Let us denote by $C^p(\W)$ the set of differential $p$-forms $\omega_p$, given in terms of local coordinates by
\begin{equation}
\omega_p=\frac{1}{p!}\omega_{\mu_1\dots\mu_p}(x)dx^{\mu_1}\wedge\dots\wedge dx^{\mu_p},
\end{equation}
where $\omega_{\mu_0\mu_1\dots\mu_p}(x)$ are scalar fields of the $(D+1)$-dimensional space-time $\W$. We also define the (co)-boundary mapping
\begin{equation}
d:C^p(\W)\rightarrow C^{p+1}(\W),
\end{equation}
as
\begin{equation}
d\omega_p=\frac{1}{p!}\partial_\alpha \omega_{\mu_1\dots\mu_p}(x)dx^\alpha\wedge dx^{\mu_1}\wedge\dots\wedge dx^{\mu_p}.
\end{equation}
We define as {\it closed} forms those for which $d\omega_p=0$, while forms written as $\omega_p=d\sigma_{p-1}$ are  defined as {\it exact}. Since the mapping $d$ is nilpotent, i.e., $d^2\omega_p=0$ for all forms, all exact forms are closed.
This differential mapping defines a long sequence among the sets of differential forms
 \be
 \begin{tikzcd}
   0\arrow{r}{d}& C^0\arrow{r}{d} &C^1 \arrow{r}{d}&\cdots\arrow{r}{d}&C^D \arrow{r}{d} & C^{D+1} \arrow{r}{d}& 0,
   \end{tikzcd}
 \ee
where the last map follows from the fact that $\W$ has a finite dimension of $(D+1)$, and the first map is the inclusion of the zero element. Hence, all $(D+1)$-forms are closed.\\

Let us now define the set of closed forms as $Z^p(\W)$ and the set of exact forms as $B^p(\W)$. Define the $p$th {\it de Rham cohomology } group by
\begin{equation}
\H^p_{dR}(\W)=Z^p(\W)/B^p(\W),
\end{equation}
by the equivalence relation among two $p$-forms $\omega_p$ and $\omega_p'$ by $\omega_p \sim \omega_p'+d\sigma_{p-1}$. Notice that the zero element in $\H^p_{dR}(\W)$ consists of all exact $p$-forms. \\

At this point, it is necessary to enumerate some important assumptions we are making throughout these lectures:
\begin{enumerate}
\item
The space-time $\W$ has a boundary, this is $\partial\W=\Sigma^D$.
\item
There is an isomorphism between $\H^\bullet_{dR}(\W)$ and the singular cohomology $\H^\bullet (\W; \R)$. Therefore, from now on, we shall refer to the singular cohomology.
\item
By the Universal Coefficient Theorem, we have that for the $p$th cohomology group
\be
\H^q(\W; \R)=\left(\H^q(\W;\Z)\otimes_{\Z}\R\right)\oplus\text{Tor}\left(\H^{q-1}(\W;\Z),\R\right),
\ee
where we shall restrict the free component to $\Z^k$ (for some $k\in\mathbb{N}$), mainly because this component will be related to quantum charges, which happen to be quantized (see the box about Dirac quantization) under a symmetry related to a continuous compact group such as $U(1)$.  This implies that 
\be
\H^q(\W;\Z)\otimes_{\Z}\R=\Z^k\otimes_{\Z}\R.
\ee
Also, for simplicity, we shall not consider the torsion in this lecture. However, the presence of torsion in cohomology plays a fundamental role in the study of more general scenarios in String Theory, and we suggest that the reader keep an eye on it.
\end{enumerate}

\subsection{Hodge dual map}
Henceforth, we are not considering the presence of torsion. This is important since, by the universal coefficient theorem, we restrict to $\H^p(\W,\Z)$ as an inclusion in $H^p(\W;\R)$. 
Under this consideration, let us now define the {\it Hodge dual} map, denoted $\ast$ by
\begin{equation}
\ast: \H^p(\W;\Z)\rightarrow \H^{D+1-p}(\W;\Z),
\end{equation}
relating each $p$-form $\omega_p$ to a $(D+1-p)$-form denoted $\ast\omega_p$ given by
\begin{equation}
\ast\omega_p=\frac{1}{p!(D+1-p)!}\epsilon^{\alpha_1\cdots\alpha_p}_{\mu_1\cdots\mu_{D+1-p}}\omega_{\alpha_1\cdots\alpha_p}dx^{\mu_1}\wedge\cdots\wedge dx^{\mu_{D+1-p}}.
\end{equation}
where $\epsilon$ is the usual Levi-Civita tensor. Notice that for any $p$-form and for any $p$, $d(\omega_p\wedge\ast\omega_p)=0$ for which $\omega_p\wedge\ast\omega_p$ defines an equivalence class in $\H^{D+1}(\W;\Z)$.\\

\subsection{Homology}
Define the set of $p$-chain\footnote{A $p$-chain  is a linear combination of $p$-dimensional simplices.} $\Sigma^p$ in $\W$ by $C_p(\W)$. Define also the boundary map as
\be
\partial : C_n\rightarrow C_{n-1},
\ee
where $\partial^2=0$. This defines the chain complex
 \be
 \begin{tikzcd}
   0\arrow{r}{\partial}& C_{D+1}\arrow{r}{\partial} &C_D \arrow{r}{\partial}&\cdots\arrow{r}{\partial}&C_p \arrow{r}{\partial} & C_{p-1} \arrow{r}{\partial}& \cdots,
   \end{tikzcd}
 \ee
 Define now a $p$-cycle as a chain without boundary; this is $\partial\Sigma^p=0$. A trivial cycle is a cycle with boundary; this is $\Sigma^p=\partial \Pi^{p+1}$.
We define {\it the $p$th homology group} as
\be
\H_p(\W;\Z)=\{ \Sigma^p\in C_p~|\partial\Sigma^p=0\}/\{\Sigma^p\in C_p~|\Sigma^p=\partial\Pi^{p+1}\}.
\ee
Hence, the $p$th homology group consists of equivalence classes of cycles, where the null element is chains that are the boundary of others.\\

Since the chains $C_n$ have a vector space structure, we can associate a dual space $C_n^\ast$ given by
\be
C_n^\ast=\text{Hom}(\Phi_n, C_n),
\ee
with $\Phi_n: C_n\rightarrow \Z$. It is a known result that $C_n^\ast\cong C^n$, establishing a way to define an isomorphism between $\H^p(\W;\Z)$ and $\H_p(\W;\Z)$.

\subsection{De Rham Theorem's}
In order to construct the isomorphism between cohomology and homology, we enunciate the following theorems due to de Rham:\\

\begin{enumerate}
\item
For $\omega_p\in C^p(\W)$ with $\omega_p=d\alpha_{p-1}$, (i.e. $[\omega_p]=[0]\in \H^p(\W;\Z)$)and $[\Sigma^p]\in \H_p(\W;\Z)$, it follows from Gauss's theorem that
\be
\int_{\Sigma^p} \omega_p=\int_{\partial\Sigma^p}\alpha_{p-1}=0.
\ee

\item
For $[\omega_p]\in\H^p(\W;\Z)$ and $\Sigma^p\in C_p(\W)$ with $\Sigma^p=\partial\Pi^{p+1}$,
\be
\int_{\Sigma^p}\omega_p=\int_{\Pi^{p+1}}d\omega_p =0.
\ee
\end{enumerate}

Due to these lemmas, we can construct an isomorphism as
\be
\Phi:\H_p(\W;\Z)\times \H^p(\W;\Z) \rightarrow \Z
\ee
such that
\be
\Phi([\Sigma^p], [\omega_p])=\int_{\Sigma^p} \omega_p \in \Z,
\ee
 or equivalently, by Hodge star,
 \be
 \Phi([\Sigma^{D+1-p}], [\ast\omega_p])=\int_{\Sigma^{D+1-p}}\ast\omega_p\in \Z.
 \label{iso}
 \ee
For our purposes, we shall use the latest expression.  Notice that  $\Phi$ is uniquely defined by equivalence classes on $\H_p$ and $\H^p$. The homeomorphism $\Phi$ is identically zero for $[\Sigma^p]=0\in\H_p(\W;\Z)$ and/or $[\omega_p]=0\in\H^p(\W;\Z)$.\\

\subsection{Poincar\'e dual map.} 
Let $\omega_p$ be a $p$-form such that $[\omega_p]\in\H^p(\W;\Z)$ and let $\Sigma^p$ be a $p$-cycle such that $[\Sigma^p]\in\H_p(\W;\Z)$. Define the {\it Poincar\'e Dual} map $\PD$ as
\be
\PD :\H_p(\W;\Z)\longleftrightarrow \H^{D+1-p}(\W;\Z),
\ee
such that $\PD (\omega_p)$ has compact support on a transversal cycle to $\Sigma^p$. This is
\be
\int_{\Sigma_{p}}\omega_p=\int_{\W} \omega_p\wedge\PD(\Sigma_p)\in \Z.
\ee
Notice that in this case
 $\PD(\Sigma_p)$ is identified with a localized bump form (usually expressed in physics literature as a Dirac's delta) supported on $\PD(\omega_p)$
   with
 \be
 \int_{\PD(\omega_p)}\PD(\Sigma^p)=1.
 \ee
 
  A similar expression is
 \be
 \int_{\Sigma^{D+1-p}}\ast\omega_p=\int_{\W} \ast\omega_p\wedge\PD(\Sigma^{D+1-p}).
 \ee
 Notice that we can always express the latest form as $\PD(\Sigma^{D+1-p})=\ast \sigma_p$. Therefore, the above integral represents an intersection between $\Sigma^{D+1-p}$ and a current formed by $\ast\sigma_p$.\\

\bibliographystyle{JHEP}
\bibliography{references.bib}

@book{peskin2018introduction,
  title={An introduction to quantum field theory},
  author={Peskin, Michael E},
  year={2018},
  publisher={CRC press}
}

@book{ryder1996quantum,
  title={Quantum field theory},
  author={Ryder, Lewis H},
  year={1996},
  publisher={Cambridge university press}
}

@book{zee2010quantum,
  title={Quantum field theory in a nutshell},
  author={Zee, Anthony},
  volume={7},
  year={2010},
  publisher={Princeton university press}
}

@book{burgess2020introduction,
  title={Introduction to effective field theory},
  author={Burgess, Cliff Peter},
  year={2020},
  publisher={Cambridge University Press}
}

@book{weinberg1995quantum,
  title={The quantum theory of fields},
  author={Weinberg, Steven},
  volume={2},
  year={1995},
  publisher={Cambridge university press}
}

@book{hatcher2005algebraic,
  title={Algebraic topology},
  author={Hatcher, Allen},
  year={2005},
  publisher={Cambridge U. Press}
}

@book{bott1982differential,
  title={Differential forms in algebraic topology},
  author={Bott, Raoul and Tu, Loring W and others},
  volume={82},
  year={1982},
  publisher={Springer}
}

@book{guillemin2019differential,
  title={Differential forms},
  author={Guillemin, Victor and Haine, Peter},
  year={2019},
  publisher={World Scientific}
}

@article{miller1994notes,
  title={Notes on cobordism},
  author={Miller, Haynes},
  journal={Notes typed by Dan Christensen and Gerd Laures based on lectures of Haynes Miller},
  year={1994}
}

@book{lawson2016spin,
  title={Spin Geometry (PMS-38), Volume 38},
  author={Lawson, H Blaine and Michelsohn, Marie-Louise},
  volume={20},
  year={2016},
  publisher={Princeton university press}
}

@book{nakahara2018geometry,
  title={Geometry, topology and physics},
  author={Nakahara, Mikio},
  year={2018},
  publisher={CRC press}
}

@book{nash1988topology,
  title={Topology and geometry for physicists},
  author={Nash, Charles and Sen, Siddhartha},
  year={1988},
  publisher={Elsevier}
}

@book{nash1991differential,
  title={Differential topology and quantum field theory},
  author={Nash, Charles},
  year={1991},
  publisher={Elsevier}
}

@book{naber1997topology,
  title={Topology, geometry, and gauge fields},
  author={Naber, Gregory L and Naber, Gregory L},
  year={1997},
  publisher={Springer}
}

@book{bertlmann2000anomalies,
  title={Anomalies in quantum field theory},
  author={Bertlmann, Reinhold A},
  volume={91},
  year={2000},
  publisher={Oxford university press}
}

@book{von2009differential,
  title={Differential forms in mathematical physics},
  author={Von Westenholz, Cornelius},
  year={2009},
  publisher={Elsevier}
}

@article{Banks:2010zn,
    author = "Banks, Tom and Seiberg, Nathan",
    title = "{Symmetries and Strings in Field Theory and Gravity}",
    eprint = "1011.5120",
    archivePrefix = "arXiv",
    primaryClass = "hep-th",
    doi = "10.1103/PhysRevD.83.084019",
    journal = "Phys. Rev. D",
    volume = "83",
    pages = "084019",
    year = "2011"
}

@article{Gaiotto:2014kfa,
    author = "Gaiotto, Davide and Kapustin, Anton and Seiberg, Nathan and Willett, Brian",
    title = "{Generalized Global Symmetries}",
    eprint = "1412.5148",
    archivePrefix = "arXiv",
    primaryClass = "hep-th",
    doi = "10.1007/JHEP02(2015)172",
    journal = "JHEP",
    volume = "02",
    pages = "172",
    year = "2015"
}

@inproceedings{Luo:2023ive,
    author = "Luo, Ran and Wang, Qing-Rui and Wang, Yi-Nan",
    title = "{Lecture Notes on Generalized Symmetries and Applications}",
    booktitle = "arXiv preprint",
    eprint = "2307.09215",
    archivePrefix = "arXiv",
    primaryClass = "hep-th",
    month = "7",
    year = "2023"
}

@article{Bhardwaj:2023kri,
    author = "Bhardwaj, Lakshya and Bottini, Lea E. and Fraser-Taliente, Ludovic and Gladden, Liam and Gould, Dewi S. W. and Platschorre, Arthur and Tillim, Hannah",
    title = "{Lectures on generalized symmetries}",
    eprint = "2307.07547",
    archivePrefix = "arXiv",
    primaryClass = "hep-th",
    doi = "10.1016/j.physrep.2023.11.002",
    journal = "Phys. Rept.",
    volume = "1051",
    pages = "1--87",
    year = "2024"
}

@article{Apruzzi:2023uma,
    author = "Apruzzi, Fabio and Bonetti, Federico and Gould, Dewi S. W. and Schafer-Nameki, Sakura",
    title = "{Aspects of Categorical Symmetries from Branes: SymTFTs and Generalized Charges}",
    eprint = "2306.16405",
    archivePrefix = "arXiv",
    primaryClass = "hep-th",
    month = "6",
    year = "2023"
}

@article{Brennan:2023mmt,
    author = "Brennan, T. Daniel and Hong, Sungwoo",
    title = "{Introduction to Generalized Global Symmetries in QFT and Particle Physics}",
    eprint = "2306.00912",
    archivePrefix = "arXiv",
    primaryClass = "hep-ph",
    month = "6",
    year = "2023"
}

@article{Schafer-Nameki:2023jdn,
    author = "Schafer-Nameki, Sakura",
    title = "{ICTP lectures on (non-)invertible generalized symmetries}",
    eprint = "2305.18296",
    archivePrefix = "arXiv",
    primaryClass = "hep-th",
    doi = "10.1016/j.physrep.2024.01.007",
    journal = "Phys. Rept.",
    volume = "1063",
    pages = "1--55",
    year = "2024"
}

@article{Bhardwaj:2023ayw,
    author = "Bhardwaj, Lakshya and Schafer-Nameki, Sakura",
    title = "{Generalized Charges, Part II: Non-Invertible Symmetries and the Symmetry TFT}",
    eprint = "2305.17159",
    archivePrefix = "arXiv",
    primaryClass = "hep-th",
    month = "5",
    year = "2023"
}

@article{Benedetti:2023ipt,
    author = "Benedetti, Valentin and Bueno, Pablo and Magan, Javier M.",
    title = "{Generalized Symmetries for Generalized Gravitons}",
    eprint = "2305.13361",
    archivePrefix = "arXiv",
    primaryClass = "hep-th",
    doi = "10.1103/PhysRevLett.131.111603",
    journal = "Phys. Rev. Lett.",
    volume = "131",
    number = "11",
    pages = "111603",
    year = "2023"
}

@article{Bhardwaj:2023wzd,
    author = "Bhardwaj, Lakshya and Schafer-Nameki, Sakura",
    title = "{Generalized Charges, Part I: Invertible Symmetries and Higher Representations}",
    eprint = "2304.02660",
    archivePrefix = "arXiv",
    primaryClass = "hep-th",
    month = "4",
    year = "2023"
}

@article{Gomes:2023ahz,
    author = "Gomes, Pedro R. S.",
    title = "{An introduction to higher-form symmetries}",
    eprint = "2303.01817",
    archivePrefix = "arXiv",
    primaryClass = "hep-th",
    doi = "10.21468/SciPostPhysLectNotes.74",
    journal = "SciPost Phys. Lect. Notes",
    volume = "74",
    pages = "1",
    year = "2023"
}

@inproceedings{Cordova:2022ruw,
    author = "Cordova, Clay and Dumitrescu, Thomas T. and Intriligator, Kenneth and Shao, Shu-Heng",
    title = "{Snowmass White Paper: Generalized Symmetries in Quantum Field Theory and Beyond}",
    booktitle = "{Snowmass 2021}",
    eprint = "2205.09545",
    archivePrefix = "arXiv",
    primaryClass = "hep-th",
    month = "5",
    year = "2022"
}

@article{Benedetti:2022zbb,
    author = "Benedetti, Valentin and Casini, Horacio and Magan, Javier M.",
    title = "{Generalized symmetries and Noether\textquoteright{}s theorem in QFT}",
    eprint = "2205.03412",
    archivePrefix = "arXiv",
    primaryClass = "hep-th",
    doi = "10.1007/JHEP08(2022)304",
    journal = "JHEP",
    volume = "08",
    pages = "304",
    year = "2022"
}

@article{Sharpe:2015mja,
    author = "Sharpe, Eric",
    title = "{Notes on generalized global symmetries in QFT}",
    eprint = "1508.04770",
    archivePrefix = "arXiv",
    primaryClass = "hep-th",
    doi = "10.1002/prop.201500048",
    journal = "Fortsch. Phys.",
    volume = "63",
    pages = "659--682",
    year = "2015"
}

@article{Palti:2019pca,
    author = "Palti, Eran",
    title = "{The Swampland: Introduction and Review}",
    eprint = "1903.06239",
    archivePrefix = "arXiv",
    primaryClass = "hep-th",
    reportNumber = "MPP-2019-53",
    doi = "10.1002/prop.201900037",
    journal = "Fortsch. Phys.",
    volume = "67",
    number = "6",
    pages = "1900037",
    year = "2019"
}

@article{vanBeest:2021lhn,
    author = "van Beest, Marieke and Calder\'on-Infante, Jos\'e and Mirfendereski, Delaram and Valenzuela, Irene",
    title = "{Lectures on the Swampland Program in String Compactifications}",
    eprint = "2102.01111",
    archivePrefix = "arXiv",
    primaryClass = "hep-th",
    doi = "10.1016/j.physrep.2022.09.002",
    journal = "Phys. Rept.",
    volume = "989",
    pages = "1--50",
    year = "2022"
}

@article{Grana:2021zvf,
    author = "Gra\~na, Mariana and Herr\'aez, Alvaro",
    title = "{The Swampland Conjectures: A Bridge from Quantum Gravity to Particle Physics}",
    eprint = "2107.00087",
    archivePrefix = "arXiv",
    primaryClass = "hep-th",
    doi = "10.3390/universe7080273",
    journal = "Universe",
    volume = "7",
    number = "8",
    pages = "273",
    year = "2021"
}

@article{Agmon:2022thq,
    author = "Agmon, Nathan Benjamin and Bedroya, Alek and Kang, Monica Jinwoo and Vafa, Cumrun",
    title = "{Lectures on the string landscape and the Swampland}",
    eprint = "2212.06187",
    archivePrefix = "arXiv",
    primaryClass = "hep-th",
    month = "12",
    year = "2022"
}

@phdthesis{Makridou:2023wkb,
    author = "Makridou, Andriana",
    title = "{Swampland program: cobordism, tadpoles, and the dark dimension}",
    doi = "10.5282/edoc.32315",
    school = "Munich U., Munich, Max Planck Inst.",
    month = "6",
    year = "2023"
}

@article{Blumenhagen:2022bvh,
    author = "Blumenhagen, Ralph and Cribiori, Niccol\`o and Kneissl, Christian and Makridou, Andriana",
    title = "{Dimensional Reduction of Cobordism and K-theory}",
    eprint = "2208.01656",
    archivePrefix = "arXiv",
    primaryClass = "hep-th",
    reportNumber = "MPP-2022-95",
    doi = "10.1007/JHEP03(2023)181",
    journal = "JHEP",
    volume = "03",
    pages = "181",
    year = "2023"
}

@article{Blumenhagen:2022mqw,
    author = "Blumenhagen, Ralph and Cribiori, Niccol\`o and Kneissl, Christian and Makridou, Andriana",
    title = "{Dynamical cobordism of a domain wall and its companion defect 7-brane}",
    eprint = "2205.09782",
    archivePrefix = "arXiv",
    primaryClass = "hep-th",
    reportNumber = "MPP-2022-57",
    doi = "10.1007/JHEP08(2022)204",
    journal = "JHEP",
    volume = "08",
    pages = "204",
    year = "2022"
}

@article{Andriot:2022mri,
    author = "Andriot, David and Carqueville, Nils and Cribiori, Niccol\`o",
    title = "{Looking for structure in the cobordism conjecture}",
    eprint = "2204.00021",
    archivePrefix = "arXiv",
    primaryClass = "hep-th",
    reportNumber = "MPP-2022-11",
    doi = "10.21468/SciPostPhys.13.3.071",
    journal = "SciPost Phys.",
    volume = "13",
    number = "3",
    pages = "071",
    year = "2022"
}

@article{Blumenhagen:2021nmi,
    author = "Blumenhagen, Ralph and Cribiori, Niccol\`o",
    title = "{Open-closed correspondence of K-theory and cobordism}",
    eprint = "2112.07678",
    archivePrefix = "arXiv",
    primaryClass = "hep-th",
    reportNumber = "MPP-2021-201",
    doi = "10.1007/JHEP08(2022)037",
    journal = "JHEP",
    volume = "08",
    pages = "037",
    year = "2022"
}

@article{Dierigl:2020lai,
    author = "Dierigl, Markus and Heckman, Jonathan J.",
    title = "{Swampland cobordism conjecture and non-Abelian duality groups}",
    eprint = "2012.00013",
    archivePrefix = "arXiv",
    primaryClass = "hep-th",
    doi = "10.1103/PhysRevD.103.066006",
    journal = "Phys. Rev. D",
    volume = "103",
    number = "6",
    pages = "066006",
    year = "2021"
}

@article{Montero:2020icj,
    author = "Montero, Miguel and Vafa, Cumrun",
    title = "{Cobordism Conjecture, Anomalies, and the String Lamppost Principle}",
    eprint = "2008.11729",
    archivePrefix = "arXiv",
    primaryClass = "hep-th",
    doi = "10.1007/JHEP01(2021)063",
    journal = "JHEP",
    volume = "01",
    pages = "063",
    year = "2021"
}

@article{McNamara:2019rup,
    author = "McNamara, Jacob and Vafa, Cumrun",
    title = "{Cobordism Classes and the Swampland}",
    eprint = "1909.10355",
    archivePrefix = "arXiv",
    primaryClass = "hep-th",
    month = "9",
    year = "2019"
}

@article{Minasian:1997mm,
    author = "Minasian, Ruben and Moore, Gregory W.",
    title = "{K theory and Ramond-Ramond charge}",
    eprint = "hep-th/9710230",
    archivePrefix = "arXiv",
    reportNumber = "YCTP-P21-97",
    doi = "10.1088/1126-6708/1997/11/002",
    journal = "JHEP",
    volume = "11",
    pages = "002",
    year = "1997"
}

@article{Witten:1998cd,
    author = "Witten, Edward",
    title = "{D-branes and K-theory}",
    eprint = "hep-th/9810188",
    archivePrefix = "arXiv",
    reportNumber = "IASSNS-HEP-98-82",
    doi = "10.1088/1126-6708/1998/12/019",
    journal = "JHEP",
    volume = "12",
    pages = "019",
    year = "1998"
}

@article{Bouwknegt:2000qt,
    author = "Bouwknegt, P. and Mathai, V.",
    title = "{D-branes, B fields and twisted K theory}",
    eprint = "hep-th/0002023",
    archivePrefix = "arXiv",
    reportNumber = "ADP-00-07-M88",
    doi = "10.1088/1126-6708/2000/03/007",
    journal = "JHEP",
    volume = "03",
    pages = "007",
    year = "2000"
}

@article{Diaconescu:2000wz,
    author = "Diaconescu, Duiliu-Emanuel and Moore, Gregory W. and Witten, Edward",
    title = "{A Derivation of K theory from M theory}",
    eprint = "hep-th/0005091",
    archivePrefix = "arXiv",
    reportNumber = "IASSNS-HEP-00-38",
    month = "5",
    year = "2000"
}

@article{Diaconescu:2000wy,
    author = "Diaconescu, Duiliu-Emanuel and Moore, Gregory W. and Witten, Edward",
    title = "{E(8) gauge theory, and a derivation of K theory from M theory}",
    eprint = "hep-th/0005090",
    archivePrefix = "arXiv",
    reportNumber = "IASSNS-HEP-00-39",
    doi = "10.4310/ATMP.2002.v6.n6.a2",
    journal = "Adv. Theor. Math. Phys.",
    volume = "6",
    pages = "1031--1134",
    year = "2003"
}

@article{Witten:2000cn,
    author = "Witten, Edward",
    editor = "Duff, Michael J. and Liu, J. T. and Lu, J.",
    title = "{Overview of K theory applied to strings}",
    eprint = "hep-th/0007175",
    archivePrefix = "arXiv",
    doi = "10.1142/S0217751X01003822",
    journal = "Int. J. Mod. Phys. A",
    volume = "16",
    pages = "693--706",
    year = "2001"
}

@article{Maldacena:2001xj,
    author = "Maldacena, Juan Martin and Moore, Gregory W. and Seiberg, Nathan",
    title = "{D-brane instantons and K theory charges}",
    eprint = "hep-th/0108100",
    archivePrefix = "arXiv",
    reportNumber = "RUNHETC-2001-24",
    doi = "10.1088/1126-6708/2001/11/062",
    journal = "JHEP",
    volume = "11",
    pages = "062",
    year = "2001"
}

@inproceedings{Moore:2003vf,
    author = "Moore, Gregory W.",
    title = "{K theory from a physical perspective}",
    booktitle = "{Symposium on Topology, Geometry and Quantum Field Theory (Segalfest)}",
    eprint = "hep-th/0304018",
    archivePrefix = "arXiv",
    reportNumber = "RUNHETC-2003-02",
    pages = "194--234",
    month = "4",
    year = "2003"
}

@article{Evslin:2006cj,
    author = "Evslin, Jarah",
    title = "{What does(n't) K-theory classify?}",
    eprint = "hep-th/0610328",
    archivePrefix = "arXiv",
    month = "10",
    year = "2006"
}

@article{Grady:2019man,
    author = "Grady, Daniel and Sati, Hisham",
    title = "{Ramond\textendash{}Ramond fields and twisted differential K-theory}",
    eprint = "1903.08843",
    archivePrefix = "arXiv",
    primaryClass = "hep-th",
    doi = "10.4310/ATMP.2022.v26.n5.a2",
    journal = "Adv. Theor. Math. Phys.",
    volume = "26",
    number = "5",
    pages = "1097--1155",
    year = "2022"
}

@article{Damian:2019bkb,
    author = "Damian, Cesar and Loaiza-Brito, Oscar",
    title = "{Some remarks on Swampland conjectures, fluxes and K-theory in IIB toroidal compactifications}",
    eprint = "1906.08766",
    archivePrefix = "arXiv",
    primaryClass = "hep-th",
    doi = "10.1016/j.aop.2023.169334",
    journal = "Annals Phys.",
    volume = "454",
    pages = "169334",
    year = "2023"
}

@article{Garcia-Compean:2013sla,
    author = "Garc\'\i{}a-Compe\'an, Hugo and Loaiza-Brito, Oscar and Mart\'\i{}nez-Merino, Aldo and Santos-Silva, Roberto",
    title = "{Half-flat Quantum Hair}",
    eprint = "1310.4557",
    archivePrefix = "arXiv",
    primaryClass = "hep-th",
    doi = "10.1103/PhysRevD.89.044025",
    journal = "Phys. Rev. D",
    volume = "89",
    number = "4",
    pages = "044025",
    year = "2014"
}

@article{Loaiza-Brito:2007umh,
    author = "Loaiza-Brito, Oscar and Oda, Kin-ya",
    title = "{Effects of brane-flux transition on black holes in string theory}",
    eprint = "hep-th/0703033",
    archivePrefix = "arXiv",
    reportNumber = "CINVESTAV-MTY-07-025, RIKEN-TH-92",
    doi = "10.1088/1126-6708/2007/08/002",
    journal = "JHEP",
    volume = "08",
    pages = "002",
    year = "2007"
}

@article{Loaiza-Brito:2003ivs,
    author = "Loaiza-Brito, Oscar",
    title = "{The Exchange of orientifold two planes in M theory}",
    eprint = "hep-th/0303270",
    archivePrefix = "arXiv",
    reportNumber = "CTP-MIT-3356",
    doi = "10.1103/PhysRevD.69.106003",
    journal = "Phys. Rev. D",
    volume = "69",
    pages = "106003",
    year = "2004"
}

@article{Loaiza-Brito:2004ajy,
    author = "Loaiza-Brito, O.",
    title = "{Instantonic branes, Atiyah-Hirzebruch spectral sequence, and SL(2,Z) duality of N=4 SYM}",
    doi = "10.1016/j.nuclphysb.2003.12.029",
    journal = "Nucl. Phys. B",
    volume = "680",
    pages = "271--301",
    year = "2004"
}

\end{document}